\documentclass[aps,prb,twocolumn,showpacs,superscriptaddress,floatfix,amsmath,amssymb,longbibliography]{revtex4-2}

\usepackage[titletoc,toc,title]{appendix}
\usepackage{graphicx}
\usepackage{bm}
\usepackage{xcolor}
\usepackage{booktabs}
\usepackage{diagbox}
\usepackage{braket}
\usepackage{amsmath}
\usepackage{setspace}
\usepackage[sort&compress]{natbib}
\usepackage[none]{hyphenat}
\usepackage{hyperref}

\renewcommand{\vec}[1]{\bm{#1}}

\begin{document}

\title{Machine learning assisted determination of electronic correlations from magnetic resonance}

\author{Anantha Rao}
\affiliation{Department of Physics, Indian Institute of Science Education and Research, Pune, Maharastra, 411008, India}
\author{Stephen Carr}
\affiliation{Department of Physics, Brown University, Providence, Rhode Island 02912-1843, USA}
\affiliation{Brown Theoretical Physics Center, Brown University, Providence, Rhode Island 02912-1843, USA.}
\author{Charles Snider}
\affiliation{Department of Physics, Brown University, Providence, Rhode Island 02912-1843, USA}
\author{D. E. Feldman}
\affiliation{Department of Physics, Brown University, Providence, Rhode Island 02912-1843, USA}
\affiliation{Brown Theoretical Physics Center, Brown University, Providence, Rhode Island 02912-1843, USA.}
\author{Chandrasekhar Ramanathan}
\affiliation{Department of Physics and Astronomy, Dartmouth College, Hanover, NH 03755, USA}
\author{V. F. Mitrovi\'{c}}
\affiliation{Department of Physics, Brown University, Providence, Rhode Island 02912-1843, USA}

\date{\today}

\begin{abstract}

In the presence of strong electronic spin correlations, the hyperfine interaction imparts long-range coupling between nuclear spins.
Efficient protocols for the extraction of such complex information about electron correlations via magnetic response are not well known.
Here, we study how machine learning can extract material parameters and help interpret magnetic response experiments.
A low-dimensional representation that classifies the strength and range of the interaction is discovered by unsupervised learning.
Supervised learning generates models that predict the spatial extent of electronic correlations and the total interaction strength.
Our work demonstrates the utility of artificial intelligence in the development of new probes of quantum systems, with applications to experimental studies of strongly correlated materials.

 
\end{abstract}

\maketitle

\section{Introduction}

Quantum phases of electronic matter, such as superconductors and spin liquids, have promising applications, but they are challenging to study due to their fragile nature~\cite{norman2021fragile, semeghini2021probing}.
Luckily, the hyperfine interaction offers a non-intrusive coupling of the electronic spin structure to nuclear spins, providing an alternate probing mechanism of the electronic phase.
As nuclear magnetic resonance (NMR) probes the nuclear spins with low-frequency pulses relative to typical electronic energies, measurement of the nuclear magnetization should not disturb the electronic ground state a priori.
Most NMR techniques in correlated systems measure how short-range electronic spin susceptibility changes the spin dynamics through dephasing, energy dissipation, or spectral (Knight) shifts (e.g. $T_2$, $T_1$, and $K$)~\cite{PhysRevB.44.12537,Rigamonti98, HorvaticSolit96, BERTHIER2017331, Mitrovic2002, Mitrovic2010,Alloul,Vinograd:2021aa,Jansa:2018aa}.

\begin{figure}
    \centering
    \includegraphics[width=\linewidth]{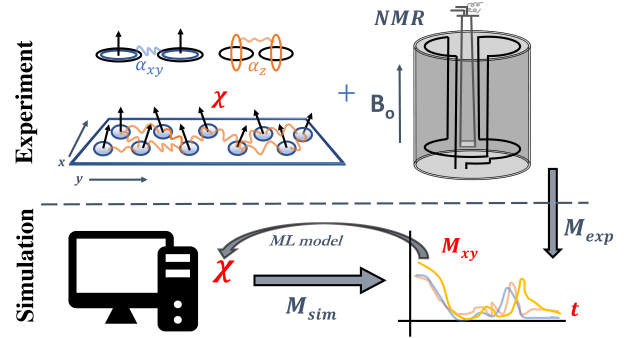}
    \caption{\textbf{A schematic for material discovery via NMR} Top: A magnetic resonance experiment in a fixed field $B_0$ allows for the controlled evolution of many nuclei, with interactions between spins mediated by the electron susceptibility $\chi$. This susceptibility is sometimes strongly polarized along either the out-of-plane ($\alpha_z$) or in-plane ($\alpha_{xy}$) spin axes. Bottom: Magnetization time-series data is generated by spin-echo simulations, represented by the map $M_{sim}: \chi \to M_{xy}$, and a machine learning (ML) model is trained to predict  the parameters of $\chi$ from them.}
    \label{nmr_setup}
\end{figure}

Measurement of decay processes has its limitations however.
Magnetic relaxation becomes complicated in frustrated materials with a highly degenerate energy spectra, and when repeated measurements heat the sample or the echo fails to satisfy time-reversal symmetry~\cite{Ishida2019, Pustogow2019, Vinograd2021, Mitrovic2008}.
Some of these confounding features have been recently reproduced by large-scale simulations of thousands of nuclear spins with long-range interactions~\cite{carr2022signatures}.
As the details of the nuclear interaction are inherited from the electronic spin susceptibility, aspects of the electronic spin-spin correlations that were previously inaccessible to NMR can be estimated after appropriate analysis.
The most successful approach required careful pulse sweeping to capture the anisotropy of the electronic spin-spin correlations.
But no diagnostic tool was developed to distinguish between exponential (short-range) or power law (long-range) decay of the interaction strength in space, although it was noted that the two types of decay produced different magnetization responses.

Here, we revisit large lattice simulations of interacting nuclear spins and ask a direct question: how much information about the electronic correlations can be extracted from a single NMR experiment on an unknown material?
Although we will consider only traditional spin-echo protocols at ideal pulsing, this work develops a framework for testing the efficacy of magnetic probes of electronic features.
After applying standard machine-learning techniques to a large dataset of time-series simulations~\cite{li2021deep, cordova2021bayesian, cobas2020nmr}, if the initial parameters are predicted at a rate better than random guessing, one can surmise that some amount of information is obtained by the proposed experiment.
Moreover, a guide for interpreting real measurements is developed from feature ranking of the simulated data~\cite{bratko1997machine, carleo2019machine}. 
In this case, we compare the  ``automatically'' generated approach to the previous analytical treatment~\cite{carr2022signatures}, and highlight improvements that the data-driven approach provides.

In Fig.~\ref{nmr_setup}, we provide a conceptual outline of this methodology, illustrating both an interacting spin system and the inverse problem of material discovery. Our work is organized as follows: In Section \ref{sec:methods}, we discuss the methods employed for simulations and learning the electronic susceptibility. In Section \ref{sec:results}, we use machine learning applications to see what information about the material is accessible from the echo measurements, and generate predictive models. Finally, in Section \ref{sec:Conlcusion}, we discuss the results in the greater context of magnetic measurements of strongly correlated systems.

\section{Methods}
\label{sec:methods}
\subsection{NMR simulations}

The combined nuclei-electron system can be understood by considering a Hamiltonian of the form
\begin{equation}
    H = H_e + H_n + H_{hf}
\end{equation}
where $H_e$ and $H_n$ are the Hamiltonians of the electrons and nuclei, respectively, and $H_{hf}$ is the interaction between the two species, mediated by the hyperfine interaction.
As we are interested in just the nuclear time-dynamics (NMR probes do not operate at frequencies relevant to electrons), we integrate out the degrees of freedom associated with the electrons to obtain a Hamiltonian that consists of just the nuclei and an electron-mediated nuclear-nuclear interaction.
We then apply a mean-field approximation to the nuclear spins, replacing direct spin-spin interactions with a mean magnetization on each nuclei $\vec{M}_i$, to arrive at~\cite{carr2022signatures}:

\begin{equation}
\label{eq:H_MF}
    H_{mf}(i) = -\nu_i I_i^z - \sum_{d=x,y,z} \alpha_d I_i^d M_i^d
\end{equation}
where the nuclei are labeled by index $i$, $I_i^d$ is the spin operator along the $d$-axis for nuclei $i$, and $\nu_i$ are the (Zeeman) precession frequencies of the nuclei. The second term represents the mean-field interaction, with $\alpha_d$ the effective electron-mediated coupling strength along nuclear spin axis $d$.
Note that the $\alpha_d$ depend on both the hyperfine interaction tensor and the electronic susceptibility tensor.
$M_i^d$ is the effective local magnetic field felt by nuclei $i$ along axis $d$ due to the mean-field interaction.
The overall strength of the electronic susceptibility is encoded by the $\alpha$ variables, while the spatial structure of the susceptibility enters into the formula for $M^d_i$.
The effective local magnetization for a nucleus at site $r_i$ is defined as the sum:
\begin{equation}
    \vec{M}_i = \frac{1}{\kappa} \sum_j K(r_{ij}) \braket{\vec{I}_j}
\end{equation}
with $K$ the ``kernel'' for the interaction that encodes the spatial structure of the electronic spin correlations, $r_{ij} \equiv r_j - r_i$ the distance to nuclei $j$, and $\kappa = \sum_j K(r_{0j})$ is a normalization constant.
We also introduce an effective length-scale for the kernel, $L$, given by the weighted average of $r$:
\begin{equation}
L = \frac{1}{\kappa} \sum_j r_{0j} K(r_{0j}).
\end{equation}

\begin{figure*}
    \centering
    \includegraphics[width=\linewidth]{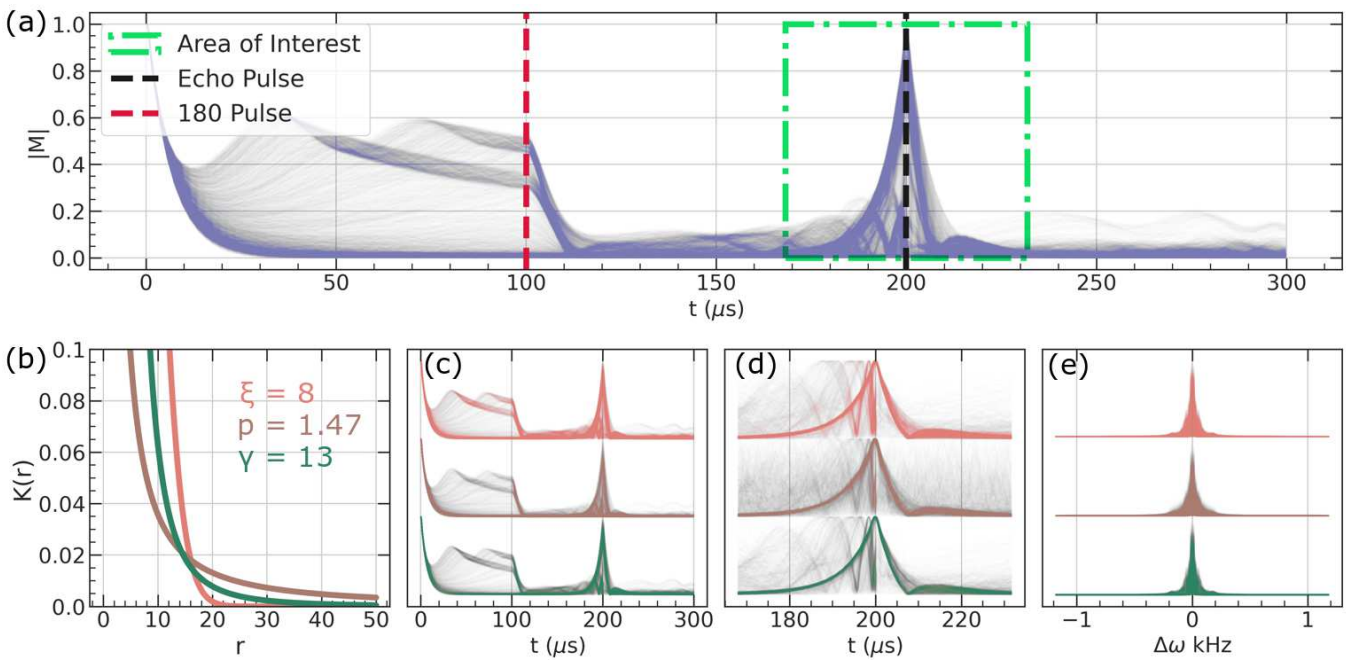}
    \caption{\textbf{Overview of the spin-echo dataset}
    (a) The echo responses over all 15,000 simulations. The refocused echo due to spin realignment is observed at $t= \tau = 200$~$\mu$s. The region around the echo, $t = [167, 233]~\mu$s, is highlighted as an ``Area of Interest'' that will serve as a truncated dataset alongside the full time-series.
    (b) Absolute value of the three magnetization kernels $K$: Gaussian (pink) with $\xi$ = 8, power (brown) with $p$ = 1.47, and RKKY (green) with $\gamma$ = 13. 
    (c) Time domain echo responses for the Gaussian, power-law and RKKY radial kernel forms.
    (d) Time domain responses within the area of interest.
    (e) Spin echo responses in frequency domain, relative to the resonance frequency.
    }
    \label{fig:spin_echo_dataset_methods}
\end{figure*}

We study three forms of $K$.
First, a short-range Gaussian that depends on a correlation length $\xi$, $K(r) = e^{-(r/\xi)^2}$, motivated by the susceptibility expected from a gapped spin excitation.
Second, a long-range form given by a power $p$, $K(r) = r^{-p}$, motivated by a gapless spin excitation.
Finally, the RKKY form expected from electron-mediated spin-spin interactions in a simple metal which is also dependent on a length $\gamma$, $K(x) = x^{-4} (x \cos{x} - \sin{x})$ for $x = 2(r/\gamma)$. 
In Fig.~\ref{fig:spin_echo_dataset_methods}b, the three functional forms for K are plotted using parameters that yield similar length scales for comparison.
We note that the Gaussian form decays rapidly, while the alternating sign in the tail of the RKKY form leads to an effective cancellation of any net magnetization far away from the nuclei.
In contrast, the power law has a characteristic long-range tail.
One of the most important parameters to extract in this model is the total effective action of the net magnetization on a single nuclear spin.
This is given by an integral of the interaction $\alpha$ over the entire lattice, which we call the ``weight'' $W$.
As we have already normalized $K$, this is simply given by: 
\begin{equation}
W = \alpha_x + \alpha_y + \alpha_z
\end{equation}

The time-dependent magnetization $M(t)$ is then obtained by averaging over all nuclei in a large interacting two-dimensional lattice. 
The time-dynamics of the spins are calculated from Eq.~\ref{eq:H_MF} using a massively parallel GPU code~\cite{carr2022signatures}.
The code implements discretized time-propagation of a density matrix following the Lindbladian formulation of quantum dynamics.
We performed calculations on a $100 \times 100$ square lattice of spin-$\tfrac{1}{2}$ nuclei, with a time step $dt = 160$~ns, delay time $\tau = 100$~$\mu$s, and with precession frequencies $\nu_i$ sampled randomly from a Lorentzian distribution with a FWHM (full width at half maximum) linewidth of 8~kHz.
We sampled $\alpha_i$ and the kernel-specific variables $\{\xi,p,\gamma \}$ such that the kernel integral $W$ lies in the range $[0.1,0.3]$ MHz.
This range of $W$ is chosen to avoid echos that have no interaction-induced features ($W < 0.1$) and cases were the echos are completely dephased (e.g. decay to zero) at our chosen $\tau$ value ($W > 0.3$).
Specifically, $\alpha_z$ and $\alpha_x = \alpha_y$ were sampled randomly in a range of $[0.01,0.1]$ MHz, $\xi \in [8,32]$, $p \in [1,3]$, and $\gamma \in [13,52]$.
A random sample was simulated if the associated $W$ was in the range $[0.1,0.3]$, otherwise that random parameter selection was discarded and a new selection was generated.
This process was looped until three datasets consisting of $5,000$ samples each were obtained for the three kernel types ($15,000$ simulations in total).
For additional details of the simulation and spin-echo protocol, see App.~\ref{app:sims}.

\subsection{Machine Learning}

We have used a combination of supervised and unsupervised learning techniques to look for patterns within the simulated time-dependent magnetization that reflect the material properties.
We employed two unsupervised learning techniques that perform dimensionality reduction, principal component analysis (PCA) and variational auto-encoders (VAEs).
Without any reference to the initial material parameters, PCA identifies which subspace of the time-series basis provides the best metric for distinguishing the response curves~\cite{jolliffe2016principal}.
VAEs offer a more sophisticated, non-linear approach to identifying important subspaces, but are more computational expensive and prone to overfitting~\cite{kingma2013auto}.
Essentially, the VAE is trained to reproduce echo responses, and the center of the network yields a two-dimensional representation for each echo, called a latent space.
One can also pick a point in this two-dimensional latent space, and then use just the decoder-side of the VAE to simulate how the magnetization response depends on the discovered latent space. 


Within a supervised setting, our primary dataset was used to train models to predict one of five characteristic properties of each curve: the strength of the interaction along the plane ($\alpha_x = \alpha_y$), the strength perpendicular to the plane ($\alpha_z$), the form of the radial kernel (Gaussian, power-law, or RKKY), the integral of the interaction over the entire lattice ($W$) and the average length-scale of the interaction over the lattice ($L$), a dimensionless quantity).
This last parameter, $L$, can be generated from the others, so although all five give important physical information, there are really only four independent properties to learn.
Models built for each predictor take the average magnetization time series from a spin-echo protocol as input and predict the property of interest.

For both supervised and unsupervised learning, each echo in the dataset is normalized such that each time-series only contains values between (0,1).
Since we are most interested in characterizing the behavior at the echo pulse, we also consider a truncated dataset which contains only the magnetic response near the spin-echo ($t \in [\frac{4}{3}\tau,\frac{7}{3}\tau$], labeled Area of Interest in Fig.~\ref{fig:spin_echo_dataset_methods}a).
The time before this window is related to the free induction decay (FID) and the effect of the $180^\circ$ pulse, and the time after this window captures only interaction-driven ringing of the echo.

To interpret the results of our supervised classification and regression models, we use two different featurization techniques.
The first is the straight-forward choice, simply use the evenly spaced magnetization values $M(t)$ within the full or truncated time-domain.
Pointwise feature ranking will reveal which time contributes the most information to the model prediction.
However, it would remain unclear if the magnetization's magnitude, slope, or curvature drives predictions.
This motivates the other choice: multiscale polynomial featurization~\cite{torrisi2020random}.
We partition the region of interest into $n$ equally-sized sections with $n=3,5$ or $10$, and fit the magnetization to a cubic-polynomial in each section:
\begin{equation}
    M(x) \sim c_0 + c_1 x + c_2 x^2 + c_3 x^3.
\end{equation}
The variable $x \equiv t - t_c$ is introduced to center the time axis at the middle of the section ($t_c$), and the coefficients of the polynomial $c_i$ in each section act as the new feature set.
The coefficients track the local constant, linear, quadratic, and cubic behaviors of the average magnetization in each section, and feature ranking highlights which of these is most critical for characterization.

\begin{figure*}
    \centering
    \includegraphics[width=\linewidth]{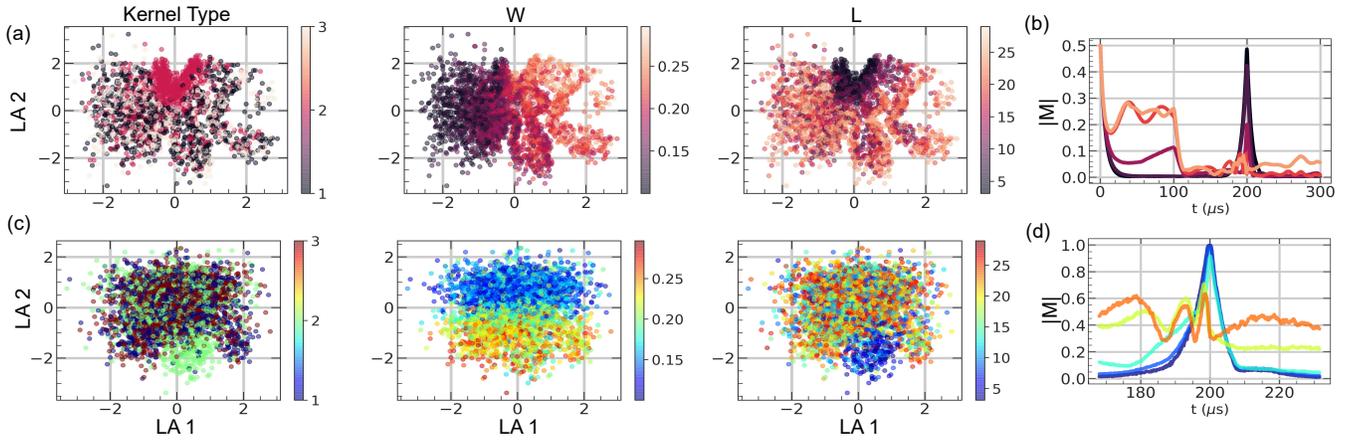}
    \caption{\textbf{Results of unsupervised learning with variational autoencoders}
    (a) The full echo response dataset, plotted in the coordinates of latent axis 1 (LA1) and latent axis 2 (LA2) obtained by the VAE.
    Each of the three columns show the same two-dimensional latent space, but with different material properties chosen to serve as the colorscale.
    These three properties are the kernel type (1: Gaussian, 2: power, 3: RKKY), the total integral of the kernel $W$, and the characteristic length-scale of the system $L$.
    (b) Absolute magnetization responses $|M(t)|$ generated by sampling the latent space along latent axis 2 (LA2).
    The sampled points are $(x,0)$, with $x$ equally spaced between $-2.5$ and $2.5$.
    (c,d) Same as (a,b), but for a VAE trained on the truncated area of interest instead of the full time-domain.
    The latent space sampling in (d) is now down along points $(0,x)$ instead.
    The colors of the lines in (b,d) correspond to the average value of $W$ at that each selected $x$ value, following the colorbar used in the $W$ column of (a,c).}
    \label{fig:unsupervised_learning_results}
\end{figure*} 

\section{Results}
\label{sec:results}

\begin{figure*}
    \centering
    \includegraphics[width=\linewidth]{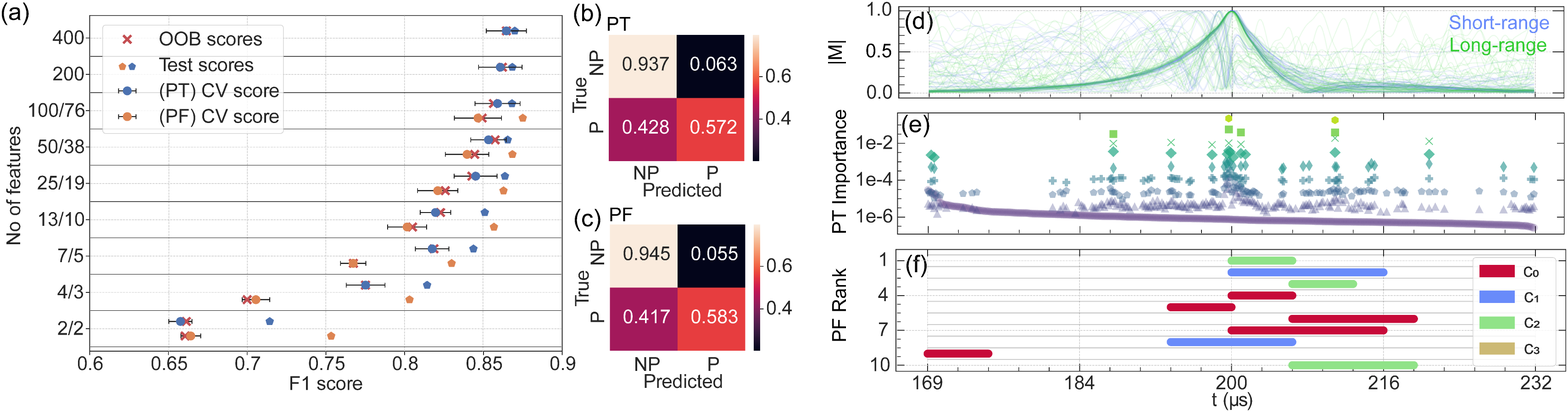}
    \caption{\textbf{Classification of the spatial extent of interaction} 
    (a) The total number of features vs $F_1$ scores of the spatial extent prediction models for both pointwise (PT) and polynomial (PF) featurization. Error bars represent the standard deviation from ten random forests trained on the same data with different initial seeds. CV: 5-fold cross-validation $F_1$ score, OOB: Out-of-bag sample scores.
    (b) The confusion matrix for a random forest model trained on all 400 PT features. The two classes are P (power) and NP (non-power), which correspond respectively to the long-range and short-range interaction kernels.
    (c) The confusion matrix for a random forest model trained on all 76 PF features.
    (d) The echo response observed in a spin-echo protocol from the short-range (NP, Gaussian and RKKY) and long-range (P, power law) kernels.
    (e) Feature importance obtained from the Gini impurity for each model of pointwise featurization. Each marker type represents the feature importance from a model trained on successively fewer features (blue: more features, green: fewer features).
    (f) Top ten ranked features from the full (76 element) polynomial featurization.
    The color of the bar indicates the coefficient order and its location and width corresponds to the time-window that it was fit to.}
    \label{fig:classification_results}
\end{figure*}

\subsection{Unsupervised learning}


We begin with unsupervised dimensionality reduction.
Such techniques are unlikely to provide any regression capability, but can generate useful classifiers for material parameters.
More importantly, if they find structure in the output which corresponds to a material parameter, it can help guide traditional analysis of the physical system.

The VAE successfully finds a useful (classifying) low-dimensional space while the PCA fails.
Although three clusters were returned by a K-nearest neighbors algorithm after PCA, these clusters showed no clear correspondence to any of the material parameters (App.~\ref{app:PCA}).
When applied to the full time-domain dataset, the two-dimensional latent-space generated by the VAE contains features that corresponded to the total interaction strength $W$, the interaction range $L$, and the kernel type (Fig.~\ref{fig:unsupervised_learning_results}(a)).
The unsupervised technique has organized simulations by increasing $W$ along its first latent space axis (LA1).
This is in agreement with the results of Ref.~\cite{carr2022signatures}, where $W$ was noted as the most important material parameter in determining the echo shape. 
Both the kernel type and the length scale $L$ were also roughly organized within the two dimensional space.
Smaller $L$ values were grouped in a $U$ shape near the point $(0,2)$, along with many of the power law kernel types.
This is partially because the sampled parameters for the power law kernel type lead to distribution of $L$ which was lower than the other two types, shown later in Fig.~\ref{fig:regression_results}(b).
Similar results are obtained after applying the VAE to only the windowed area of interest near the echo, but now the variable $W$ is ordered along LA2 instead (Fig.~\ref{fig:unsupervised_learning_results}(c)).

We ran specific points in the latent space through the decoder network, generating characteristic $M(t)$ curves in each case.
As the first axis of the latent space (LA1) corresponded more to $W$ than the second (LA2), we sample the points $(x,0)$ with $x \in [-2.5,2.5]$ in the latent space.
The echo responses at low LA1 (corresponding to small $W$) exhibit a clean echo peak while those at large LA1 (large $W$) show a suppressed peak at the echo pulse and significant ringing after time $\tau$ ($180^\circ$ pulse) and $2\tau$ (echo).
This again follows the intuition developed in the previous analytical work, and proves that unsupervised learning can generate scientifically significant interpretations of magnetic resonance measurements.
These insights are gained effectively ``for free,'' requiring only a suitably sampled dataset and access to basic machine learning software.
In situations with more material parameters, or for protocols with multiple pulse durations and axes, it can be challenging to make analytic progress due to the complicated nature of the resulting Hamiltonians.
Unsupervised learning can immediately check which key material parameters control physical observables, and yield the representative features of the response curves as that parameter is varied.

\subsection{Classification of the spatial extent of interactions}

The main shortcoming of the previous analytical work was an inability to predict if a material had short-range (exponential) or long-range (power law) decay in the nuclear coupling~\cite{carr2022signatures}.
By comparing simulations across the three different kernel forms with similar interaction weights $W$, subtle differences in the resulting $M(t)$ curves were noticed.
But for a randomly sampled material, distinguishing between kernel forms seemed impossible.
Note that this classification of short vs long-range should not be confused with the value of $L$.
Rather, short vs long describes if the interaction is exponentially localized or not.

Here, we utilize random forest models to generate a classification model for the spatial extent of the interactions: short-range vs long-range.
Since the dataset is imbalanced, we partition it into a training and test set and oversample the training set's minority class (short-range, e.g. power law)~\cite{chawla2002smote}.
By oversampling only the training data, none of the information in the test data is double counted, avoiding the introduction of synthetic observations.
While our models were unable to distinguish between the RKKY and Gaussian kernel, classification of power law vs non-power law interactions showed significantly better performance than random guessing (Fig.~\ref{fig:classification_results}(b,c)).
Such a classification is important, as a power law decay indicates that the susceptibility arises from a gapless spin system, while an exponential decay suggest the presence of a gap in the excitation spectra.
Although the RKKY type interaction arises from spin interactions in a metal (gapless), it's oscillating power-law tail on average contributes zero total magnetization.
In the mean-field simulations, this makes it act more similarly to the exponentially localized (Gaussian) interaction, but using a full many-body method may change this.

For classifying non-power law data, the two featurization methods (pointwise vs polynomial) yield nearly identical confusion matrices, but the correlation between the two models' decisions were not checked.
The predictions of non-power law simulations are nearly perfect, with very few false positives (top row of the confusion matrices).
However, for power-law simulations the decisions are only slightly better than random guessing, with a false positive rate just above 0.4 (lower row of the confusion matrices).

We contrast the performance between the pointwise and polynomial featurization schemes as a function of the number of retained features (Fig.~\ref{fig:classification_results}(a)).
To identify the essential features of the classification model, we train successive classification models that only have access to features above the median feature importance of the previous model.
This allows us to create a sequence of successively simpler models.
We continue this process for both featurization models until only two features remains, but because the pointwise data starts with more features than the polynomial model the pairings are not of identical feature count.
The models trained on pointwise features perform slightly better within each pair, but also retain roughly $30\%$ more features than the polynomial model.
Although the test scores for the two featurizations are quite similar despite this discrepancy, the 5-fold cross validation scores are significantly worse for the polynomial featurization (PF), hinting at severe over-fitting.
When we restrict the models to using only the two most important features obtained by this scheme, the polynomial model performs slightly better.

The two most critical time-value features are the magnitude of the magnetization at the echo pulse (2$\tau = 200$ $\mu$s) and the magnitude in the post-echo shoulder at approximately 210 $\mu$s.
For classifying the range of the interaction, the relative size of the echo's decay (the value at $2\tau$) to the intensity of the post-echo refocusing (shoulder) encodes key information about the range of the interaction.
The polynomial featurization ranking in Fig.~\ref{fig:classification_results}(f) shows that the slope and curvature at the echo pulse being the most important, with the mean values near the echo ranked as the fourth and fifth most important instead.
Overall, this is fairly consistent with the theoretical understanding of gained from study of the mean-field model~\cite{carr2022signatures}.
Short-range interactions led to more strongly suppressed echoes and reduced post-echo ringing, which are encoded in the slope and curvature near the echo and the post-echo shoulder.
Nevertheless we are now armed with a model that can make predictions instead of relying on pure intuition.

As mentioned previously, the most common error in our classifier were false positives for the power-law radial kernel (Fig.~\ref{fig:classification_results}(b,c)).
Motivated by the latent structure identified by the VAE, we also train separate models for different ranges of the total interaction strength $W$.
That is to say, we can refine the classification models by first using a robust prediction of $W$, as $W$ and the radial-decay type both play a role in determining the shape of $M(t)$.
We train three families of random forest models, each constrained only to consider simulations that have $W$ in the range $[0.1, 0.17]$, $[0.17, 0.23]$, or $[0.23, 0.3]$, all in units of MHz.
With this approach, the false positive error rate is reduced to about 0.2 for both the pointwise and polynomial featurization (App.~\ref{app:W_screened}), which is half the error rate when not using a $W$ filter.
As the total interaction strength $W$ explains a large amount of the $M(t)$ variations, using $W$ screening leads to better classification models of the other physical parameters.

\subsection{Prediction of the effective interaction strength}

\begin{figure}
    \centering
    \includegraphics[width=\linewidth]{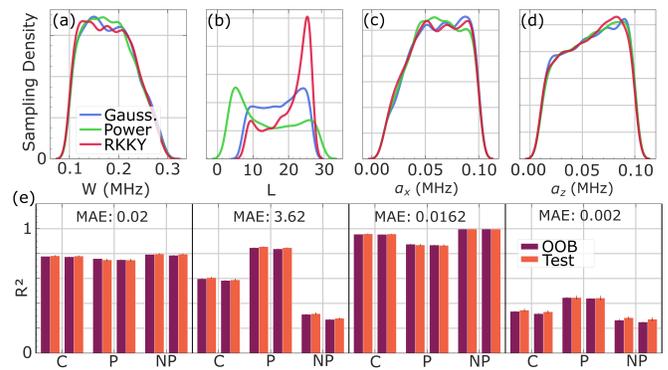}
    \caption{\textbf{Results of regression from random forest models.}
    (a-d) The distribution of the predictor variables for each kernel type.
    (e) The Out-of-Bag score and test set scores are plotted for the complete dataset (C), just power law simulations (P), and all non-power law simulations (NP). For each dataset, the first two bars represent results from the pointwise data while the last two bars are results from polynomial features. (MAE: Mean Absolute Error)}
    \label{fig:regression_results}
\end{figure}

Finally, we now discuss models that take the $M(t)$ curves as input and return predictions of the value of the material's interaction parameters, with results shown in Fig.~\ref{fig:regression_results}(e).
The prediction of $\alpha_x$ and $W$ are very accurate, with $R^2$ values of approximately $0.9$ and $0.8$, respectively.
This further justifies $W$'s use as a screening step before using a kernel-type classifier.
The models were unable to make good predictions of the out-of-plane coupling prefactor ($\alpha_z$).
This is not surprising, as our dataset consists of only simulations where the interaction-free magnetization lies entirely in-plane (due to the $90^\circ$-$180^\circ$ pulsing sequence).
When there is no out-of-plane magnetization, the $\alpha_z$ term does not affect the spin dynamics (Eq.~\ref{eq:H_MF}).
What the poor training of $\alpha_z$ confirms is that, even with significant in-plane interactions ($\alpha_x = \alpha_y$), the out-of-plane net magnetization remains very small, and therefore the value of $\alpha_z$ has little effect on resulting spin echo.
As one third of the value of $W$ is obtained from $\alpha_z$, this explains the slightly worse performance of learning $W$ compared to learning $\alpha_x$ directly.

To check for any cross-correlation between the kernel type and the four studied parameters of Fig.~\ref{fig:regression_results}, we also ran the regression models after partitioning the dataset based on the range of interaction (power ``P'' vs non-power ``NP'').
As before, we oversample the minority class (P) in the training set due to the mismatch in dataset sizes (5,000 vs 10,000).
The overall performance does not change significantly (Fig.~\ref{fig:regression_results}(e)), as predictions slightly improve for the model trained on only short-range interactions (non-power, NP) but worsen for those trained on long-range (power, P) model.
Of the four trained parameters, only $L$ shows significant deviation between the $P$ and $NP$ data-sets.
The model successfully predicts $L$ for the power law simulations, but performs poorly on the non-power law simulations.
However, the analytic approach provided \textit{no} method for the accurate prediction of $L$~\cite{carr2022signatures}.
To see an $R^2$ value above $0.8$, even when restricted to just power-law simulations, is a large improvement gained by the use of machine learning.

To extract physical intuition for the proposed regression schemes, we also analyze the feature ranking for each predictor variable in App.~\ref{app:regression_importance}
We find that the most important features for predicting $W$ are the average value ($c_0$) in the pre-echo region ($t < 2 \tau$) and the slope and curvature of the magnetization right at the echo-pulse ($c_1$, $c_2$).
We also find that the relative feature importance changes when considering just the power law or non-power law decays. 
It indicates that short-range and long-range interactions affect the magnetization response in different ways, even at similar $W$.
This encourages further development of better pulse sequences for studying materials with large electronic spin susceptibility, as even an unrefined spin-echo protocol is already capturing some scale-dependent behaviors.

\section{Conclusion}
\label{sec:Conlcusion}

We have used data-driven approaches to understand the effects of strong electronic spin correlations on nuclear spin dynamics.
By applying machine learning methods to these spin echo simulations, an effective probe of some spin susceptibility properties were developed. 
Unsupervised learning and feature-ranking of classification models provided insight into which physical parameters best describe the interacting system.
The difficult task of classifying the time-series based on the radial kernel was performed adequately by our random-forest models, with better performance achieved after sectioning the data into three bins of interaction magnitude $W$, yielding an $F_1$ score of $0.9$.

From a broader perspective, we have demonstrated that machine learning can be used to develop inference techniques when applied to magnetic resonance experiments.
Developing combined theoretical and experimental tools in this manner can improve of our understanding of real materials not only by providing better traditional probes, but also probes designed on real-time feedback between measurement and a machine-learning model~\cite{Mavadia2017,Radovic2018,Kutsukake2020}.
A simple experiment could be run to estimate the total interaction strength $W$, and then a more complicated sequence that is tailored for the resulting $W$ could be run to extract accurate information about the effective interaction length-scale $L$ and its normalized strength $\alpha_i$. 

Revisiting this problem with a deep learning algorithm may provide better predictions of $L$ and $\alpha_i$, but requires a much larger dataset.
On the other hand, designing specific pulse sequences which can access different aspects of the spin dynamics may lead to a larger improvement in our predictions.
The development of highly specialized pulse sequences in order to capture the symmetry of certain physical parameters is a well developed field for on-site or nearest-neighbor interactions~\cite{Levitt2007,Levitt2008,Schwartz2018}.
Similar protocols for systems with long-range nuclear coupling are unexplored.
Reinforcement learning could possibly be leveraged to develop such sequences.

The adoption of machine learning as an interpretable method to analyze and infer the properties of complex materials is an ongoing task. 
As evidenced by our work, it can decode some of the wealth of information obtainable from magnetic resonance experiments.

\begin{acknowledgments}
This work was supported by the National Science Foundation under grant No. OIA-1921199. 
ASR acknowledges support from the Google Summer of Code program and the KVPY fellowship by Department of Science and Technology (Govt. of India). 
The calculations were conducted using computational resources and services at the Center for Computation and Visualization, Brown University.
\end{acknowledgments}

\appendix

\section{Simulation of nuclear spins}
\label{app:sims}
To achieve a spin echo in our simulations, all spins begin in alignment along the $z$ axis, and at time $t = 0$ an $I^x$ pulse rotates the spins about the $x$ axis by an angle $\theta$.
At time $t = \tau$, an $I^y$ pulse rotates them by 2$\theta$ about the y-axis.
When $\theta = 90^\circ$, the second pulse inverts the spins and further time propagation begins to cancel any accumulated phases from the variations in each spin's resonant frequency $\nu$.
This forms a spin echo at $t = 2\tau$.
We simulate this process on a $100 \times 100$ lattice of spin-1/2 interacting nuclei~\cite{carr2022signatures} for different values of $(\alpha_x=\alpha_y, \alpha_z, K, l_K)$. $\alpha_i$ is the strength of the interaction along axis~$i = \{x,y,z\}$, $K$ is the type of radial kernel and $l_K$ is a parameter that defines the form of the specified kernel (with $l_K$ either $\xi$, $p$, or $\gamma$).

To pre-process the simulated data before usage by a machine learning algorithm, each echo-response time series is normalised such that all values of net magnetization lie between (0,1).
Since we are interested in the behavior of such systems at the echo-pulse, we also generate a dataset which consists of only a narrow time-window (of width $67$~$\mu$s) around the echo-pulse (centered at $t=200$~$\mu$s). 

\section{Unsupervised learning}
\label{app:unsupervised}
\subsection{Principal component analysis}
\label{app:PCA}
\begin{figure}
    \centering
    \includegraphics[width=\linewidth]{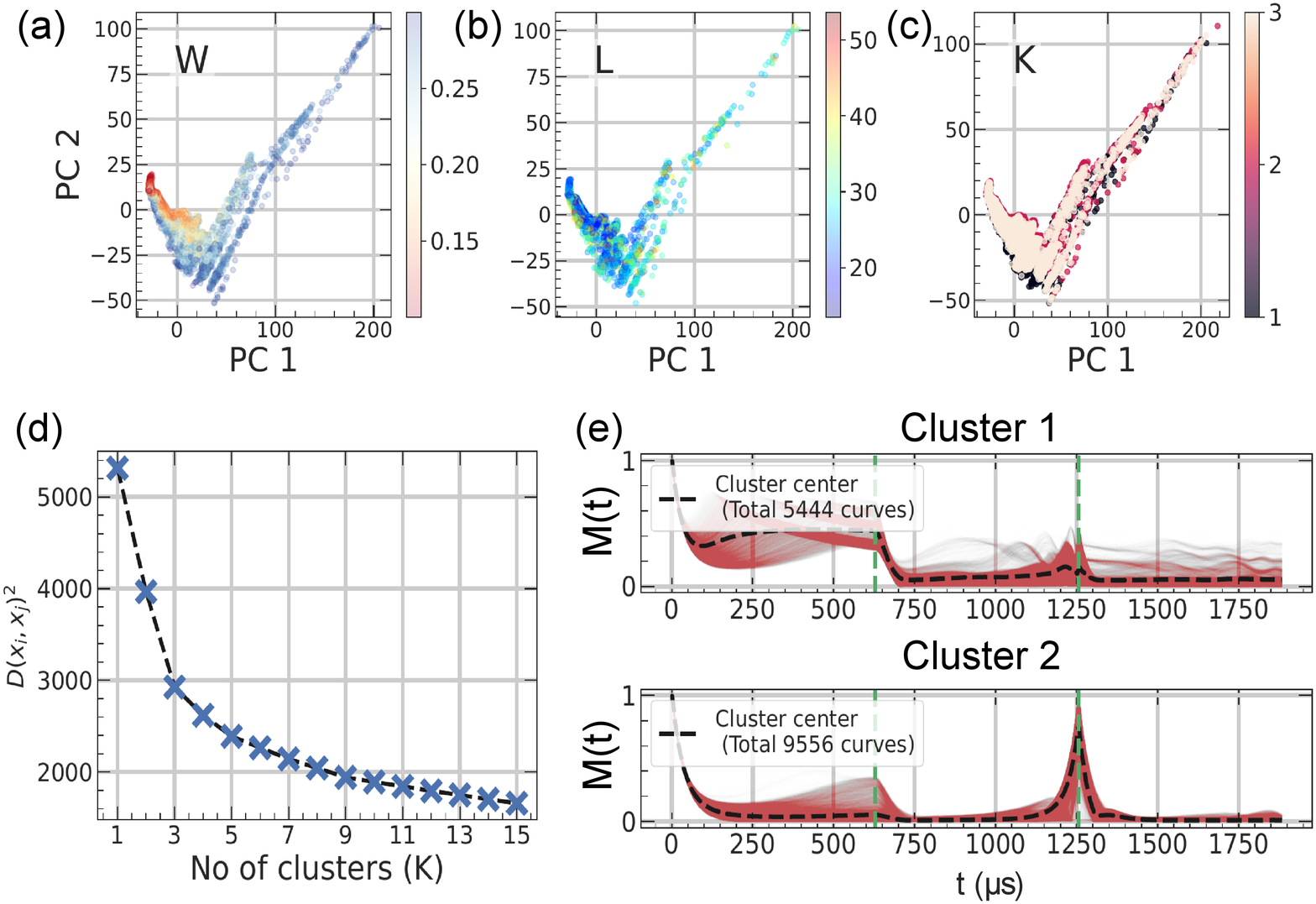}
    \caption{\textbf{PCA and K-Means results.} (a-c) The 2D PCA of the centered echo-response dataset. Each plot is colored based on the (a) total kernel integral $W$, (b) average length-scale $L$ or (c) the type of radial kernel $K$. There is no discernible difference or pattern amongst the principal components.
    (d) K-Means clustering reveals the optimum number of clusters in the echo-response must be three from the sum of squared error (SSE) scree plot 
    (e) Magnetization curves for all simulations in the two clusters, which correspond to low (high) kernel integrals $W$.}
    \label{pca_results}
\end{figure}

Principal component analysis (PCA) is a dimensionality reduction technique that generates a low-dimensional representation of a large dataset by finding the uncorrelated variables that maximize variance.
The task of identifying these new variables can be mapped to an eigenvalue problem where the principal components (eigenvectors) and their contribution to the dataset can be obtained from singular value decomposition (SVD) of the (centred) data matrix.
We used the scikit-learn package to perform PCA on our datasets. Figure \ref{pca_results} depicts a two-dimensional (2D) PCA of the echo-response data. 

\subsection{K-means clustering}
\label{app:kmeans}
K-Means clustering is an unsupervised learning technique often used to identify distinct, non-overlapping clusters in a dataset. The primary objective is to minimize the variation within each ``cluster,'' which is the mean distance between every (vectorized) datapoint assigned to that cluster.
This is achieved by minimizing the objective function:
\begin{equation}
D(K,X) = ^{\text{minimize}}_{C_1,...,C_k}\left(\sum_{k=1}^K \frac{1}{|C_k|}\sum_{i=1}^{P_k} (x_{i} - \mu_{k})^2 \right)
\end{equation}
where $K$ is the number of clusters, $x \in C_K \subset X$ are one of the $P_k$ elements of the $k$'th cluster, and $\mu_k$ is the mean value of the $k$'th cluster.
Here, we ﬁrst specify the desired number of clusters $K$, and then the K-means algorithm assigns each observation to exactly one of the $K$ clusters.

The result for a K-means cluster with varying $K$ for our dataset is given in Fig.~\ref{pca_results}d.
We find that $K=3$ gives the last useful clustering, as increasing beyond that gives little improvement in the objective. 
In Fig.~\ref{pca_results}e we show the typical magnetization curves for the two clusters for the optimized $K=2$ case, showing clear differences between cluster 1 (large $W$) and cluster 2 (small $W$).

\subsection{Variational Autoencoders}
\label{app:vae}
Variational autoencoders (VAEs)~\cite{kingma2013auto} are generative models based on layered neural networks.
Assume our dataset can be described as a set of independent and identically distributed data points, $X=\{x^{(i)}\}$ with $x^{(i)} \in R^n$.
Further, assume this data is sampled from a distribution with Gaussian distributed latent variables $z$ and model parameters $\theta$, $p_{\theta}(x^{(i)}|z)$.
Then, finding the exact posterior density $p_{\theta}(z|x^{(i)})$, e.g. solving the inverse problem of obtaining latent variables from elements of the dataset, is often an intractable problem.
VAEs approximate the true posterior distribution with a tractable approximate model $q_{\phi}(z|x^{(i)})$, with parameters $\phi$, and provide an efficient procedure to sample efficiently from $p_{\theta}(x^{(i)}|z)$.
In practice, a VAE is a network composed of three main components.
An encoder (1) projects the input into a latent space (2), and then a decoder (3) attempts to reconstruct the input from the latent representation.
After the network is trained, one can sample according to the original distribution by dropping the encoder and sampling the latent space directly.
The model is trained by minimizing (over $\theta$ and $\phi$) the cost function:
\begin{equation}
\begin{split}
J_{\theta, \phi, x^{(i)}} = &-\mathbb{E}_{z\sim q_{\phi}(z|x^{((i)})}[\log{p_{\theta}(x^{(i)}|z})] \\
&+ \beta D_{KL}(q_{\phi}(z|x^{(i)}) || p_{\theta}(z))
\end{split}
\end{equation}

The first term, the reconstruction loss consists of the expected negative log-likelihood of the $i^{\text{th}}$ data-point and favors choices of $(\theta, \phi)$ that lead to accurate reconstructions of the input. 
The second term, the regularization loss $D$, is the Kullback-Leibler divergence between the encoder’s distribution $q_{\phi}(z|x^{(i)})$ and the Gaussian prior on $z$.
A full treatment and derivations of the variational objective are given in Ref.~\onlinecite{kingma2013auto}. 

In our setup, we use a $\beta$-VAE, where $\beta \neq 1$ is an additional weight (prefactor) assigned to the KL-divergence term in the loss function.
This modifies the relative importance of the reconstruction and regularization losses.
Such $\beta$ tuning has been useful in identifying latent structures in highly non-linear datasets~\cite{higgins2016beta}.
In our simulations, we use $\beta=2$. In addition, to prevent over-fitting, we partition the dataset into training and a validation sets and train until the reconstruction loss on the validation set stops improving for 7 epochs. 
Two different VAE models are trained for the full echo-response and the time-window echo-response data.

For the full echo-response data, the encoder has the following number of nodes in successive layers: $(1882, 512, 128, 16, (2,2) )$ where (2,2) is the latent space containing the results of the encoder. The regularization term constrains this space to represent the mean and log-variance of a multivariate standardized Gaussian.
The decoder has the reversed architecture, $(2, 16, 128, 512, 1882)$, with the input to the decoder sampled from the encoder's resulting latent space via the reparametrization trick~\cite{kingma2013auto}.
All layers in both the encoder and decoder uses a LReLU (non-linear Leaky rectified linear units) activation function with the leak set to 0.01, except for the final layer which uses a sigmoid activation function.

The architecture for the echo-window data is similar, with an encoder of shape $(400, 128, 32, 16, (2,2))$, a deoder of shape $(2, 16, 32, 128, 400)$, and using the same activation functions.

\section{Supervised learning}
\label{app:supervised}
\subsection{Random forest}
\label{app:forest}
A random forest model is an ensemble technique that utilizes multiple decision trees to learn a dependent variable.
For classification, the dependent variable is determined via a majority vote from multiple trees each trained on randomly chosen (``bootstrapped'') samples and features from the training dataset.
This method of averaging over many decision trees reduces overfitting errors from a single model applied to the entire training dataset.

Within a random forest classification model, each tree is generated by iterative branching of yes/no decisions, known as a binary tree.
Eventually, the branching process terminates, and each end node is assigned a specific category (in our case, kernel type $K$).
Ideally, each branch would put all members of the same category into the same child node.
This goal is formulated as an optimization problem by minimizing the Gini impurity of each node via modification of the binary decision parameters.
The Gini impurity for a dataset with $C$ categories is defined as
\begin{equation}
1 - \sum_{j=1}^{C}p_j^2
\end{equation}
where $p_j$ represents the fraction of data which is of category $j$.
If the resulting dataset is entirely of one type (often called a ``pure'' node), for example $p_1 = 0$ and $p_2 = 1$, then the Gini impurity will be zero.
But if the dataset is exactly mixed, $p_1 = p_2 = 0.5$, then the Gini impurity will be $0.5$.

For regression, our training criteria is instead the mean squared error,
\begin{equation}
MSE = \frac{1}{N} \sum_{j=1}^N (y_i - f(x_i))^2
\end{equation}
where $y_i$ represents the label associated with the $ith$ input $x_i$ and $f$ represents the regression model.
Then, each parameter of a tree is optimized via a gradient descent algorithm to minimize the error.

\subsection{Performance metrics}
\label{app:metrics}
To measure the quality of our classification models, we adopt the $F_1$ score.
The $F_1$ score can be understood as the harmonic mean between two common classificiation metrics, the precision (Pr) and recall (Rc) scores.
All three are defined below based on the relative rates of true positives (TP), false positives (FP), and false negatives (FN) from the model: 

\begin{equation}
\begin{split}
\text{Pr} &= \frac{\text{TP}}{\text{TP} + \text{FP}} \\
\text{Rc} &= \frac{\text{TP}}{\text{TP} + \text{FN}} \\
F_1 &= 2 \times \frac{\text{Pr}\times\text{Rc}}{\text{Pr}+\text{Rc}}
\end{split}
\end{equation}

\begin{table}
  \centering
  \label{tab:supervised-classification}
  \begin{tabular}{c|l|l}
    \hline
    \textbf{Sr.No} & \textbf{Algorithm} & \textbf{$F_1$-score} \\ \hline
    1 & Logistic Regression & 0.795    \\\hline
    2 & K-nearest neighbours & 0.793   \\ \hline
    3 & Support Vector Support (SVM)  & 0.796  \\ \hline
    4 & Kernel SVM  & 0.69  \\ \hline
    5 & SGD Classifier  & 0.79  \\ \hline
    6 & DecisionTree  & 0.756  \\ \hline
    7 & Random Forests  & 0.82  \\ \hline
  \end{tabular}
  \newline\newline
  \caption{$F_1$ scores of different standard models applied to the problem of classifying the nuclear interaction as a long-range (power law) or short-range (non-power law) functional. Sr.No is the internal index for that model.}
\end{table}

To assess the quality of our regression models, we use the $R^2$ score:
\begin{equation}
    R^2 = 1 - \frac{\sum_i (y_i - f(x_i))^2}{\sum_i (y_i - \bar{y})^2}
\end{equation}
where $y_i$ are the true values, $f(x_i)$ are the predicted values, and $\bar{y}$ is the mean of all values in the training set.

\subsection{Out of Bag estimation:}
\label{app:oob}
When optimizing a random forest model, the decision trees are repeatedly fit to bootstrapped subsets of the observations such that each ``bagged'' tree uses around two-thirds of the total observations~\cite{james2013introduction}.
The remaining one-third of the observations not used to ﬁt a given bagged tree constitute the out-of-bag (OOB) observations.
We then predict the response for a given observation using each of the trees in which that observation was OOB.
This effectively allows one to treat each element of the dataset as both a training and testing element, as the OOB labeling system ensures predictions of that observation do not use any trees that were trained on it.
In order to obtain a single prediction over many trees, we average over all predicted responses (regression) or take a majority vote (classiﬁcation). 
After an OOB prediction is obtained for each element of the dataset, the overall OOB MSE or classiﬁcation error is computed.
The resulting OOB error is a valid estimate of the test error, since the response for each observation is predicted using only the trees that were not ﬁt using that observation. 
In our work, we utilize the \texttt{oob\_score} functionality in Sklearn~\cite{pedregosa2011scikit}.

\subsection{Polynomial featurization}
\label{app:poly}
\begin{figure*}
    \centering
    \includegraphics[width=\linewidth]{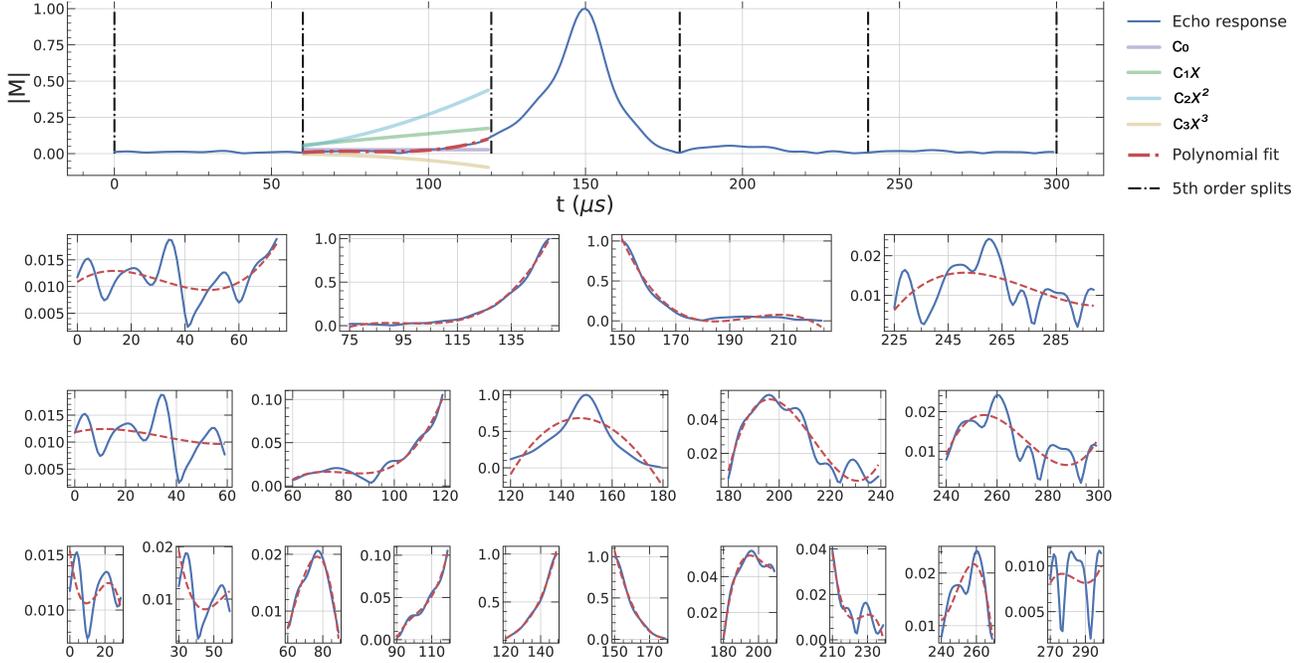}
    \caption{\textbf{Multiscale Polynomial featurization.}
    A sample echo response of three hundred pointwise features is used to perform polynomial featurization on four, five and ten equally-spaced sections.
    The coefficients of the cubic polynomial in each section of a split serve as new features for the echo response.
    The initial three hundred pointwise features can be replaced with $4 \times (4+5+10) = 76$ features.
    These features (polynomial features) are then used as input by various learning algorithms.}
    \label{fig:polyfeaturization}
\end{figure*}

To turn pointwise data into polynomial features, we simply subdivide the time-series into $N$ equal sections, and then fit a cubic polynomial $\sum_i c_i x^i$ in each section.
This process is shown for an example spin-echo simulation in Fig.~\ref{fig:polyfeaturization}.

Polynomial featurization can provide clearer physical insight into the behaviour of complex magnetic responses.
Specifically, the curvature and slope of each feature is captured in the $c_2$ and $c_1$ terms, while the mean and any non-linear or ringing behaviour is captured by the $c_0$ and $c_3$ coefficients.
Feature ranking these coefficients reveals the region and shape of the response that is most important when predicting a physical property.

\subsection{Classification with $W$ screening}
\label{app:W_screened}
\begin{figure}
    \centering
    \includegraphics[width=\linewidth]{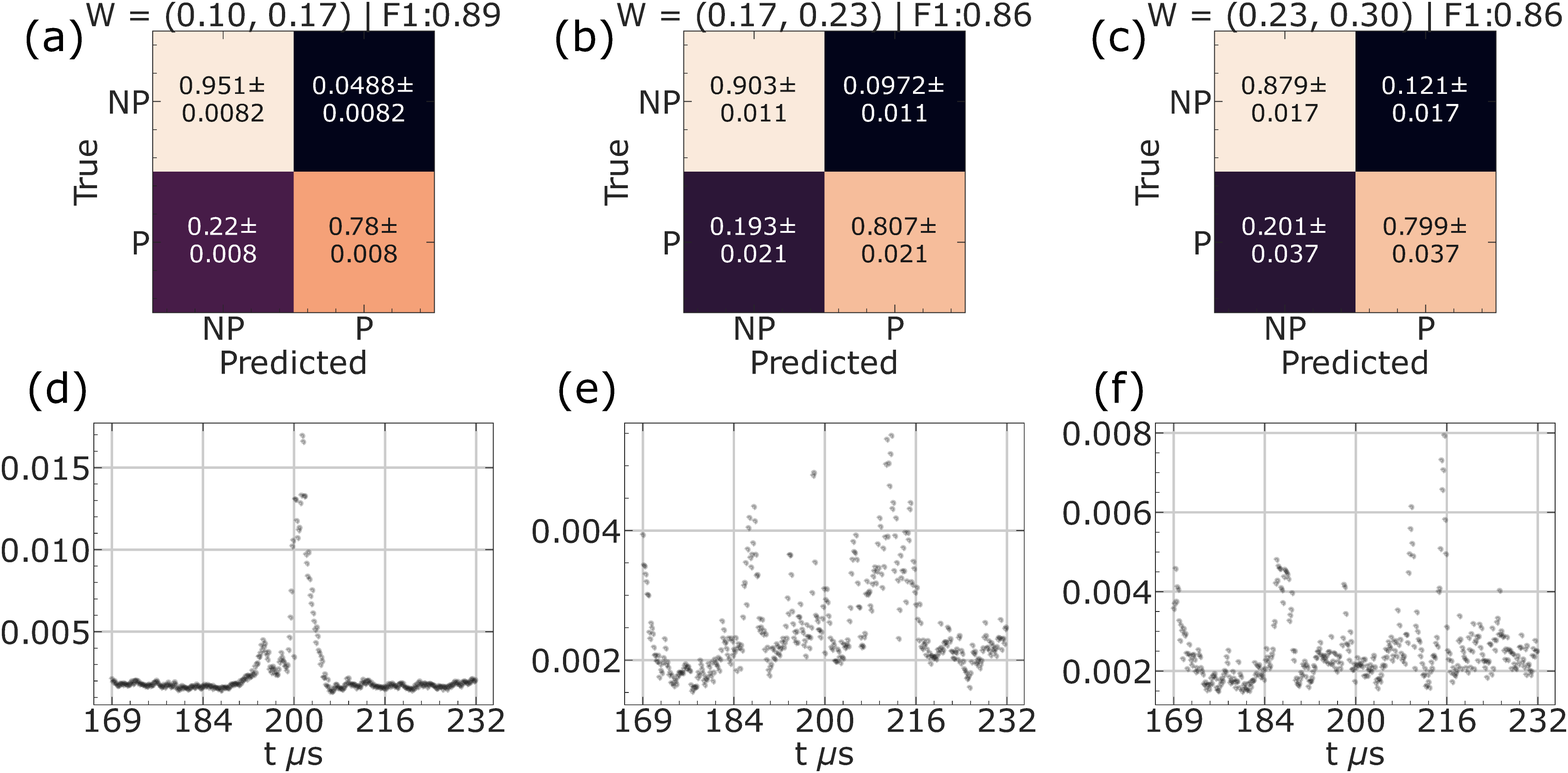}
    \caption{\textbf{Classification of radial kernel functions} based on the total value of the kernel integral from the time-series data.
    (a-c) The confusion matrix for each $W$ range.
    (d-f) Feature importance based on Gini impurity.}
    \label{fig:kernel_classification_ts}
\end{figure}

\begin{figure}
    \centering
    \includegraphics[width=\linewidth]{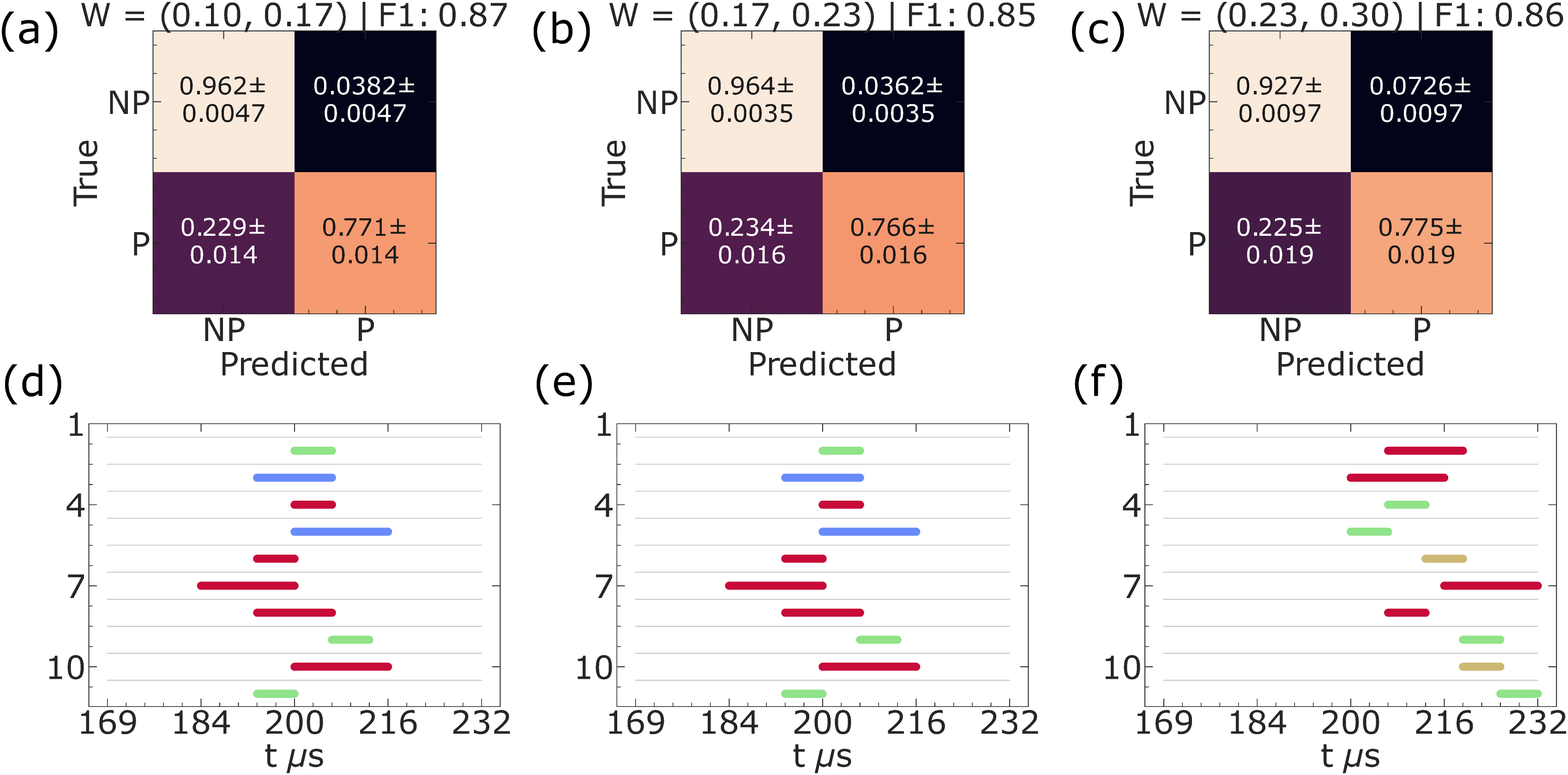}
    \caption{\textbf{Classification of radial kernel functions} based on the total value of the kernel integral from the multiscale polynomial featurization of echo-responses.
    (a-c) Confusion matrices for each bin of selected $W$.
    (d-f) The top ten features in each $W$ bin, averaged over ten models with different random seeds. The most important features are plotted at the top, with the width of the bar giving the time-window of that polynomial fit and the color giving the power of the coefficient.}
    \label{fig:kernel_classification_pf}
\end{figure}

In Fig.~\ref{fig:kernel_classification_ts} we present the confusion matrices and relative feature importance for the power-law classifier models trained within reduced ranges of the total interaction strength ($W$) value, using the time-series features.
In Fig.~\ref{fig:kernel_classification_pf} we show the same results, but now for the polynomial features.
All models significant improvement by reducing the false-error rate of power-law simulations (the results for the non-power law simulations are similar to the non-screened models).

\subsection{Feature importance for regression models}
\label{app:regression_importance}
\begin{figure}
    \centering
    \includegraphics[width=\linewidth]{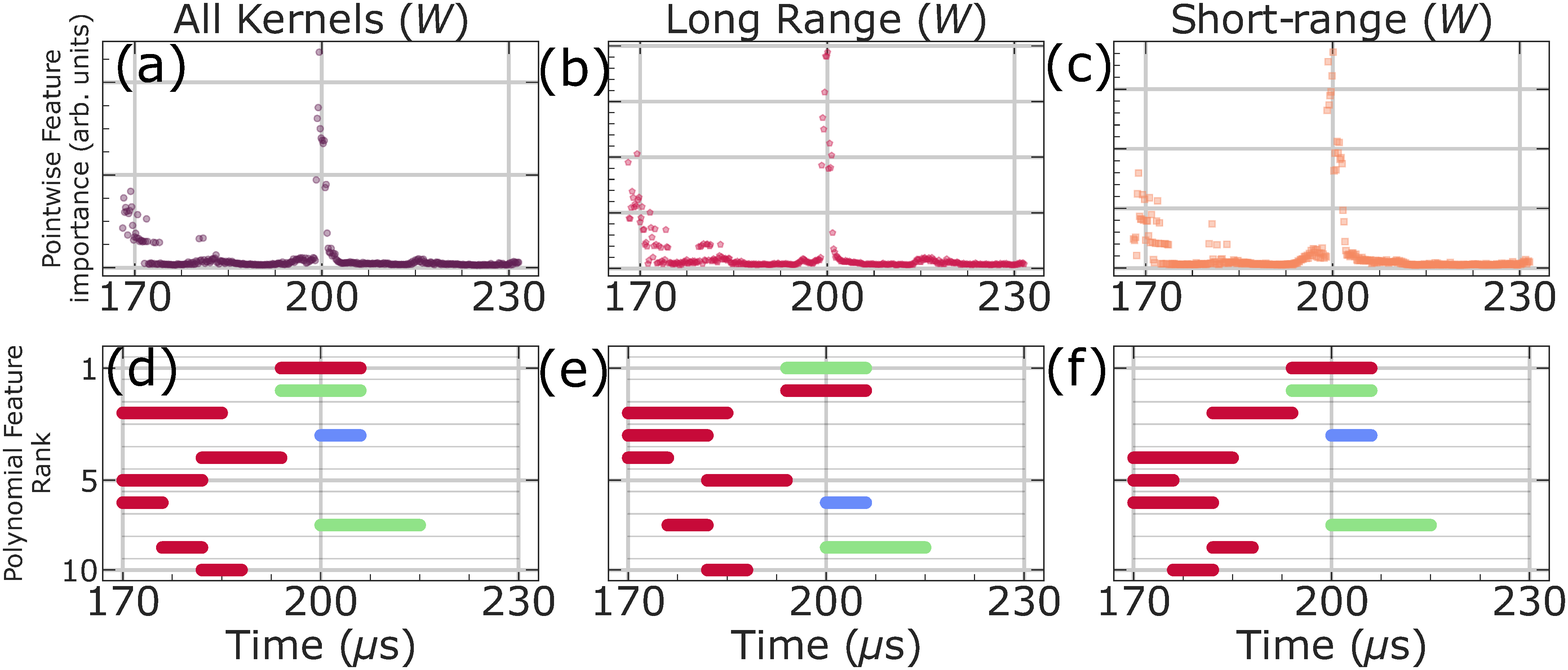}
    \caption{\textbf{Feature ranking for $W$ regression models.}
    (a-c) The median pointwise feature importances when predicting $W$ using the complete dataset, a dataset with only long-range interactions (power) and one with only short-range interactions (non-power).
    (d-f) The top ten polynomial features averaged over ten different models. Each model is trained on the same data but with a random seed.}
    \label{fig:regression_fi_results_W}
\end{figure}

\begin{figure}
    \centering
    \includegraphics[width=\linewidth]{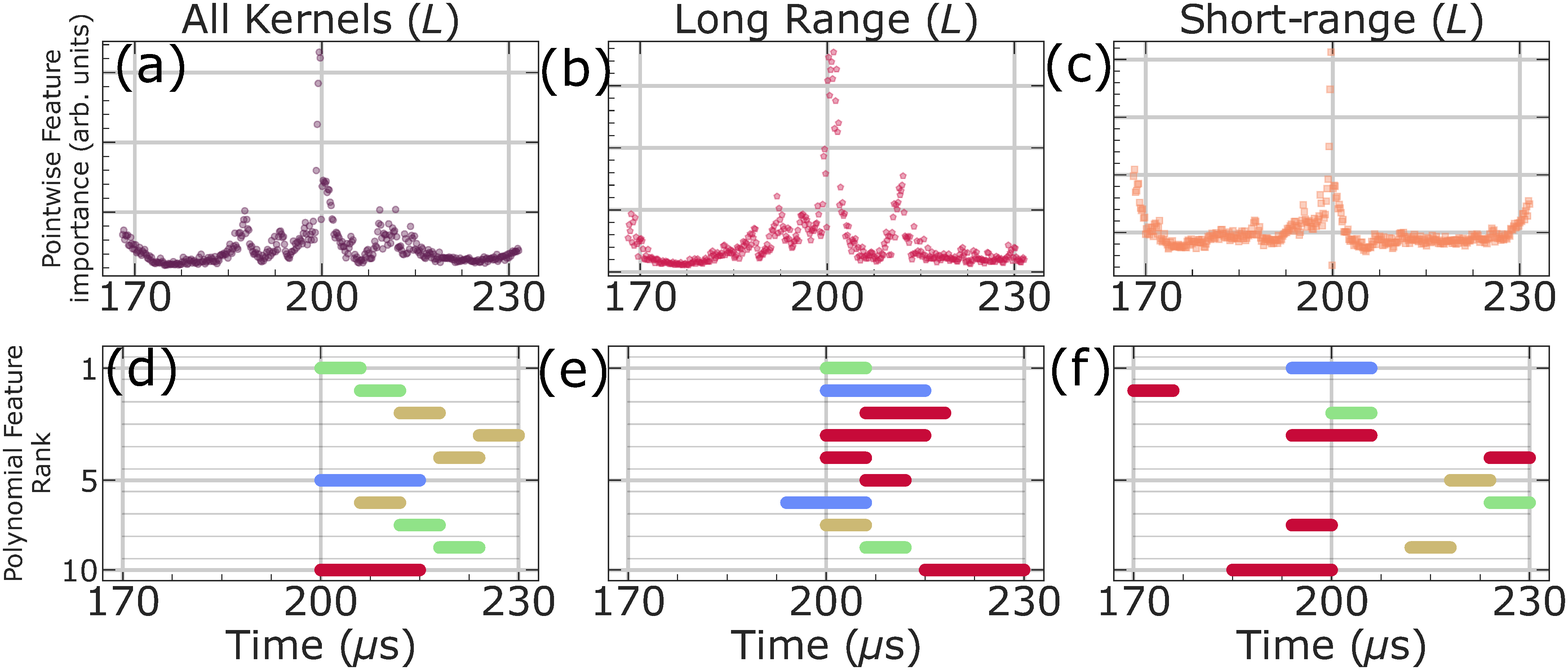}
    \caption{\textbf{Feature ranking for $L$ regression models.}
    (a-c) The median pointwise feature importances when predicting $L$ using the complete dataset, a dataset with only long-range interactions (power) and one with only short-range interactions (non-power).
    (d-f) The top ten polynomial features averaged over ten different models. Each model is trained on the same data but with a random seed.}
    \label{fig:regression_fi_results_L}
\end{figure}

\begin{figure}
    \centering
    \includegraphics[width=\linewidth]{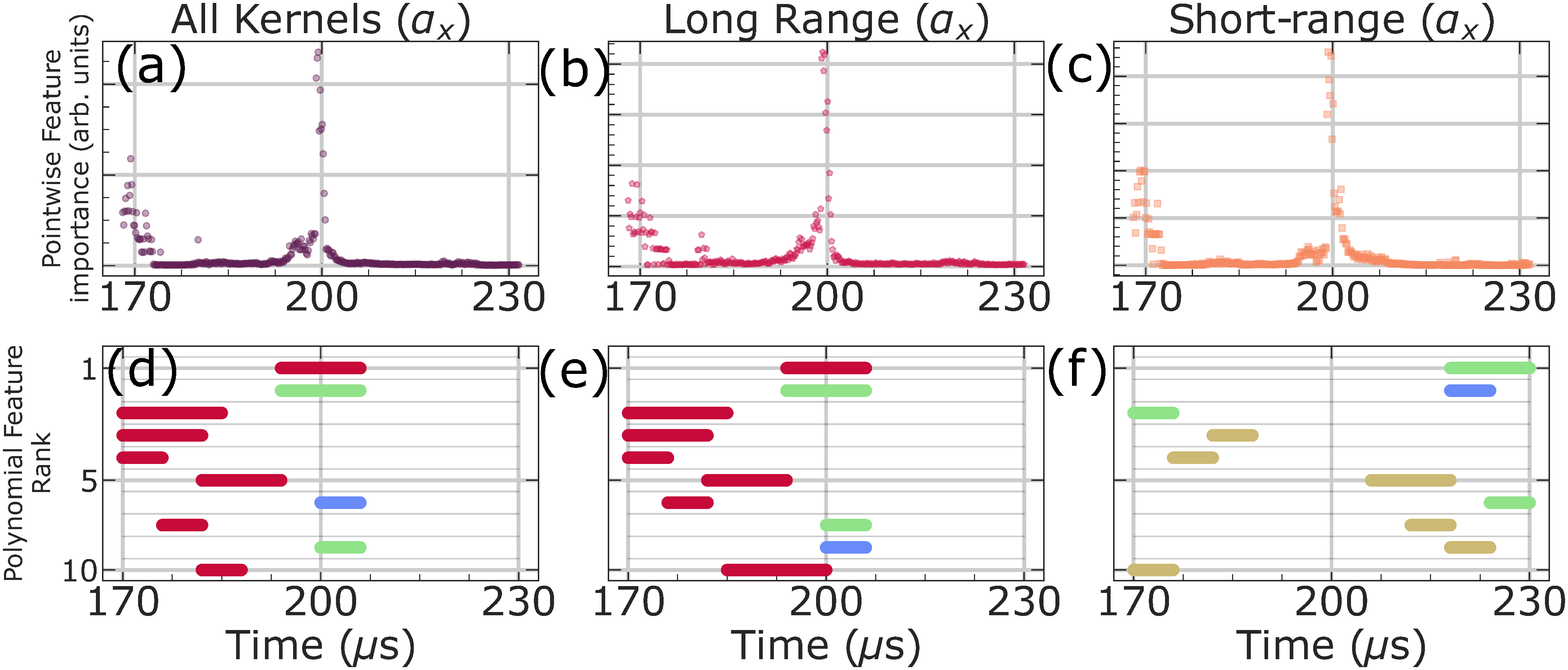}
    \caption{\textbf{Feature ranking for $\alpha_x$ regression models.}
    (a-c) The median pointwise feature importances when predicting $\alpha_x$ using the complete dataset, a dataset with only long-range interactions (power) and one with only short-range interactions (non-power).
    (d-f) The top ten polynomial features averaged over ten different models. Each model is trained on the same data but with a random seed.}
    \label{fig:regression_fi_results_alpha}
\end{figure}

In Fig. \ref{fig:regression_fi_results_W}, we present the relative feature importances of both the time-series and the polynomial featurizations for a $W$ regression model.
Included are results for the model trained on the complete dataset (``All Kernels'') as well as for models trained on only the power-law kernels (``Long Range'') and the non-power kernels (``Short Range'').
Fig.~\ref{fig:regression_fi_results_L} and Fig.~\ref{fig:regression_fi_results_alpha} show the relative feature importances for the $L$ and $\alpha_x$ regression models, respectively.
In all models, the data near the echo time $\tau = 200$~$\mu$s are the most important, with the pre-echo shoulder near $\tau = 170$~$\mu$s also of relevance in some.

\clearpage
\bibliography{refs}

\begin{thebibliography}{34}%
\makeatletter
\providecommand \@ifxundefined [1]{%
 \@ifx{#1\undefined}
}%
\providecommand \@ifnum [1]{%
 \ifnum #1\expandafter \@firstoftwo
 \else \expandafter \@secondoftwo
 \fi
}%
\providecommand \@ifx [1]{%
 \ifx #1\expandafter \@firstoftwo
 \else \expandafter \@secondoftwo
 \fi
}%
\providecommand \natexlab [1]{#1}%
\providecommand \enquote  [1]{``#1''}%
\providecommand \bibnamefont  [1]{#1}%
\providecommand \bibfnamefont [1]{#1}%
\providecommand \citenamefont [1]{#1}%
\providecommand \href@noop [0]{\@secondoftwo}%
\providecommand \href [0]{\begingroup \@sanitize@url \@href}%
\providecommand \@href[1]{\@@startlink{#1}\@@href}%
\providecommand \@@href[1]{\endgroup#1\@@endlink}%
\providecommand \@sanitize@url [0]{\catcode `\\12\catcode `\$12\catcode
  `\&12\catcode `\#12\catcode `\^12\catcode `\_12\catcode `\%12\relax}%
\providecommand \@@startlink[1]{}%
\providecommand \@@endlink[0]{}%
\providecommand \url  [0]{\begingroup\@sanitize@url \@url }%
\providecommand \@url [1]{\endgroup\@href {#1}{\urlprefix }}%
\providecommand \urlprefix  [0]{URL }%
\providecommand \Eprint [0]{\href }%
\providecommand \doibase [0]{https://doi.org/}%
\providecommand \selectlanguage [0]{\@gobble}%
\providecommand \bibinfo  [0]{\@secondoftwo}%
\providecommand \bibfield  [0]{\@secondoftwo}%
\providecommand \translation [1]{[#1]}%
\providecommand \BibitemOpen [0]{}%
\providecommand \bibitemStop [0]{}%
\providecommand \bibitemNoStop [0]{.\EOS\space}%
\providecommand \EOS [0]{\spacefactor3000\relax}%
\providecommand \BibitemShut  [1]{\csname bibitem#1\endcsname}%
\let\auto@bib@innerbib\@empty
\bibitem [{\citenamefont {Norman}(2021)}]{norman2021fragile}%
  \BibitemOpen
  \bibfield  {author} {\bibinfo {author} {\bibfnamefont {M.~R.}\ \bibnamefont
  {Norman}},\ }\bibfield  {title} {\bibinfo {title} {Fragile superconductivity
  at high magnetic fields},\ }\href {https://doi.org/10.1073/pnas.2100372118}
  {\bibfield  {journal} {\bibinfo  {journal} {Proceedings of the National
  Academy of Sciences}\ }\textbf {\bibinfo {volume} {118}},\ \bibinfo {pages}
  {e2100372118} (\bibinfo {year} {2021})}\BibitemShut {NoStop}%
\bibitem [{\citenamefont {Semeghini}\ \emph {et~al.}(2021)\citenamefont
  {Semeghini}, \citenamefont {Levine}, \citenamefont {Keesling}, \citenamefont
  {Ebadi}, \citenamefont {Wang}, \citenamefont {Bluvstein}, \citenamefont
  {Verresen}, \citenamefont {Pichler}, \citenamefont {Kalinowski},
  \citenamefont {Samajdar}, \citenamefont {Omran}, \citenamefont {Sachdev},
  \citenamefont {Vishwanath}, \citenamefont {Greiner}, \citenamefont
  {Vuleti{\'c}},\ and\ \citenamefont {Lukin}}]{semeghini2021probing}%
  \BibitemOpen
  \bibfield  {author} {\bibinfo {author} {\bibfnamefont {G.}~\bibnamefont
  {Semeghini}}, \bibinfo {author} {\bibfnamefont {H.}~\bibnamefont {Levine}},
  \bibinfo {author} {\bibfnamefont {A.}~\bibnamefont {Keesling}}, \bibinfo
  {author} {\bibfnamefont {S.}~\bibnamefont {Ebadi}}, \bibinfo {author}
  {\bibfnamefont {T.~T.}\ \bibnamefont {Wang}}, \bibinfo {author}
  {\bibfnamefont {D.}~\bibnamefont {Bluvstein}}, \bibinfo {author}
  {\bibfnamefont {R.}~\bibnamefont {Verresen}}, \bibinfo {author}
  {\bibfnamefont {H.}~\bibnamefont {Pichler}}, \bibinfo {author} {\bibfnamefont
  {M.}~\bibnamefont {Kalinowski}}, \bibinfo {author} {\bibfnamefont
  {R.}~\bibnamefont {Samajdar}}, \bibinfo {author} {\bibfnamefont
  {A.}~\bibnamefont {Omran}}, \bibinfo {author} {\bibfnamefont
  {S.}~\bibnamefont {Sachdev}}, \bibinfo {author} {\bibfnamefont
  {A.}~\bibnamefont {Vishwanath}}, \bibinfo {author} {\bibfnamefont
  {M.}~\bibnamefont {Greiner}}, \bibinfo {author} {\bibfnamefont
  {V.}~\bibnamefont {Vuleti{\'c}}},\ and\ \bibinfo {author} {\bibfnamefont
  {M.~D.}\ \bibnamefont {Lukin}},\ }\bibfield  {title} {\bibinfo {title}
  {Probing topological spin liquids on a programmable quantum simulator},\
  }\href {https://doi.org/10.1126/science.abi8794} {\bibfield  {journal}
  {\bibinfo  {journal} {Science}\ }\textbf {\bibinfo {volume} {374}},\ \bibinfo
  {pages} {1242} (\bibinfo {year} {2021})}\BibitemShut {NoStop}%
\bibitem [{\citenamefont {Ioffe}\ \emph {et~al.}(1991)\citenamefont {Ioffe},
  \citenamefont {Kivelson},\ and\ \citenamefont {Larkin}}]{PhysRevB.44.12537}%
  \BibitemOpen
  \bibfield  {author} {\bibinfo {author} {\bibfnamefont {L.~B.}\ \bibnamefont
  {Ioffe}}, \bibinfo {author} {\bibfnamefont {S.}~\bibnamefont {Kivelson}},\
  and\ \bibinfo {author} {\bibfnamefont {A.~I.}\ \bibnamefont {Larkin}},\
  }\bibfield  {title} {\bibinfo {title} {{Spin correlations and NMR relaxation
  rates in strongly correlated electron systems}},\ }\href
  {https://doi.org/10.1103/PhysRevB.44.12537} {\bibfield  {journal} {\bibinfo
  {journal} {Phys. Rev. B}\ }\textbf {\bibinfo {volume} {44}},\ \bibinfo
  {pages} {12537} (\bibinfo {year} {1991})}\BibitemShut {NoStop}%
\bibitem [{\citenamefont {Rigamonti}\ \emph {et~al.}(1998)\citenamefont
  {Rigamonti}, \citenamefont {Borsa},\ and\ \citenamefont
  {Carretta}}]{Rigamonti98}%
  \BibitemOpen
  \bibfield  {author} {\bibinfo {author} {\bibfnamefont {A.}~\bibnamefont
  {Rigamonti}}, \bibinfo {author} {\bibfnamefont {F.}~\bibnamefont {Borsa}},\
  and\ \bibinfo {author} {\bibfnamefont {P.}~\bibnamefont {Carretta}},\
  }\bibfield  {title} {\bibinfo {title} {{Basic aspects and main results of
  NMR-NQR spectroscopies in high-temperature superconductors}},\ }\href
  {https://doi.org/10.1088/0034-4885/61/10/002} {\bibfield  {journal} {\bibinfo
   {journal} {Reports on Progress in Physics}\ }\textbf {\bibinfo {volume}
  {61}},\ \bibinfo {pages} {1367} (\bibinfo {year} {1998})}\BibitemShut
  {NoStop}%
\bibitem [{\citenamefont {Fagot-Revurat}\ \emph {et~al.}(1996)\citenamefont
  {Fagot-Revurat}, \citenamefont {Horvati\ifmmode~\acute{c}\else \'{c}\fi{}},
  \citenamefont {Berthier}, \citenamefont {S\'egransan}, \citenamefont
  {Dhalenne},\ and\ \citenamefont {Revcolevschi}}]{HorvaticSolit96}%
  \BibitemOpen
  \bibfield  {author} {\bibinfo {author} {\bibfnamefont {Y.}~\bibnamefont
  {Fagot-Revurat}}, \bibinfo {author} {\bibfnamefont {M.}~\bibnamefont
  {Horvati\ifmmode~\acute{c}\else \'{c}\fi{}}}, \bibinfo {author}
  {\bibfnamefont {C.}~\bibnamefont {Berthier}}, \bibinfo {author}
  {\bibfnamefont {P.}~\bibnamefont {S\'egransan}}, \bibinfo {author}
  {\bibfnamefont {G.}~\bibnamefont {Dhalenne}},\ and\ \bibinfo {author}
  {\bibfnamefont {A.}~\bibnamefont {Revcolevschi}},\ }\bibfield  {title}
  {\bibinfo {title} {{NMR Evidence for a Magnetic Soliton Lattice in the
  High-Field Phase of CuGe${\mathrm{O}}_{3}$}},\ }\href
  {https://doi.org/10.1103/PhysRevLett.77.1861} {\bibfield  {journal} {\bibinfo
   {journal} {Phys. Rev. Lett.}\ }\textbf {\bibinfo {volume} {77}},\ \bibinfo
  {pages} {1861} (\bibinfo {year} {1996})}\BibitemShut {NoStop}%
\bibitem [{\citenamefont {Berthier}\ \emph {et~al.}(2017)\citenamefont
  {Berthier}, \citenamefont {Horvati{\'c}}, \citenamefont {Julien},
  \citenamefont {Mayaffre},\ and\ \citenamefont
  {Kr{\"a}mer}}]{BERTHIER2017331}%
  \BibitemOpen
  \bibfield  {author} {\bibinfo {author} {\bibfnamefont {C.}~\bibnamefont
  {Berthier}}, \bibinfo {author} {\bibfnamefont {M.}~\bibnamefont
  {Horvati{\'c}}}, \bibinfo {author} {\bibfnamefont {M.-H.}\ \bibnamefont
  {Julien}}, \bibinfo {author} {\bibfnamefont {H.}~\bibnamefont {Mayaffre}},\
  and\ \bibinfo {author} {\bibfnamefont {S.}~\bibnamefont {Kr{\"a}mer}},\
  }\bibfield  {title} {\bibinfo {title} {Nuclear magnetic resonance in high
  magnetic field: Application to condensed matter physics},\ }\href
  {https://doi.org/https://doi.org/10.1016/j.crhy.2017.09.009} {\bibfield
  {journal} {\bibinfo  {journal} {Comptes Rendus Physique}\ }\textbf {\bibinfo
  {volume} {18}},\ \bibinfo {pages} {331} (\bibinfo {year} {2017})}\BibitemShut
  {NoStop}%
\bibitem [{\citenamefont {Mitrovi\ifmmode~\acute{c}\else \'{c}\fi{}}\ \emph
  {et~al.}(2002)\citenamefont {Mitrovi\ifmmode~\acute{c}\else \'{c}\fi{}},
  \citenamefont {Bachman}, \citenamefont {Halperin}, \citenamefont {Reyes},
  \citenamefont {Kuhns},\ and\ \citenamefont {Moulton}}]{Mitrovic2002}%
  \BibitemOpen
  \bibfield  {author} {\bibinfo {author} {\bibfnamefont {V.~F.}\ \bibnamefont
  {Mitrovi\ifmmode~\acute{c}\else \'{c}\fi{}}}, \bibinfo {author}
  {\bibfnamefont {H.~N.}\ \bibnamefont {Bachman}}, \bibinfo {author}
  {\bibfnamefont {W.~P.}\ \bibnamefont {Halperin}}, \bibinfo {author}
  {\bibfnamefont {A.~P.}\ \bibnamefont {Reyes}}, \bibinfo {author}
  {\bibfnamefont {P.}~\bibnamefont {Kuhns}},\ and\ \bibinfo {author}
  {\bibfnamefont {W.~G.}\ \bibnamefont {Moulton}},\ }\bibfield  {title}
  {\bibinfo {title} {Pseudogap in
  {${\mathrm{YBa}}_{2}{\mathrm{Cu}}_{3}{\mathrm{O}}_{7\ensuremath{-}\ensuremath{\delta}}$}
  from {NMR} in high magnetic fields},\ }\href
  {https://doi.org/10.1103/PhysRevB.66.014511} {\bibfield  {journal} {\bibinfo
  {journal} {Phys. Rev. B}\ }\textbf {\bibinfo {volume} {66}},\ \bibinfo
  {pages} {014511} (\bibinfo {year} {2002})}\BibitemShut {NoStop}%
\bibitem [{\citenamefont {Koutroulakis}\ \emph {et~al.}(2010)\citenamefont
  {Koutroulakis}, \citenamefont {Stewart}, \citenamefont
  {Mitrovi\ifmmode~\acute{c}\else \'{c}\fi{}}, \citenamefont
  {Horvati\ifmmode~\acute{c}\else \'{c}\fi{}}, \citenamefont {Berthier},
  \citenamefont {Lapertot},\ and\ \citenamefont {Flouquet}}]{Mitrovic2010}%
  \BibitemOpen
  \bibfield  {author} {\bibinfo {author} {\bibfnamefont {G.}~\bibnamefont
  {Koutroulakis}}, \bibinfo {author} {\bibfnamefont {M.~D.}\ \bibnamefont
  {Stewart}}, \bibinfo {author} {\bibfnamefont {V.~F.}\ \bibnamefont
  {Mitrovi\ifmmode~\acute{c}\else \'{c}\fi{}}}, \bibinfo {author}
  {\bibfnamefont {M.}~\bibnamefont {Horvati\ifmmode~\acute{c}\else
  \'{c}\fi{}}}, \bibinfo {author} {\bibfnamefont {C.}~\bibnamefont {Berthier}},
  \bibinfo {author} {\bibfnamefont {G.}~\bibnamefont {Lapertot}},\ and\
  \bibinfo {author} {\bibfnamefont {J.}~\bibnamefont {Flouquet}},\ }\bibfield
  {title} {\bibinfo {title} {{Field Evolution of Coexisting Superconducting and
  Magnetic Orders in ${\mathrm{CeCoIn}}_{5}$}},\ }\href
  {https://doi.org/10.1103/PhysRevLett.104.087001} {\bibfield  {journal}
  {\bibinfo  {journal} {Phys. Rev. Lett.}\ }\textbf {\bibinfo {volume} {104}},\
  \bibinfo {pages} {087001} (\bibinfo {year} {2010})}\BibitemShut {NoStop}%
\bibitem [{\citenamefont {Alloul}(2015)}]{Alloul}%
  \BibitemOpen
  \bibfield  {author} {\bibinfo {author} {\bibfnamefont {H.}~\bibnamefont
  {Alloul}},\ }\bibfield  {title} {\bibinfo {title} {{NMR in strongly
  correlated materials}},\ }\href {https://doi.org/10.4249/scholarpedia.30632}
  {\bibfield  {journal} {\bibinfo  {journal} {Scholarpedia}\ }\textbf {\bibinfo
  {volume} {10}},\ \bibinfo {pages} {30632} (\bibinfo {year}
  {2015})}\BibitemShut {NoStop}%
\bibitem [{\citenamefont {Vinograd}\ \emph
  {et~al.}(2021{\natexlab{a}})\citenamefont {Vinograd}, \citenamefont {Zhou},
  \citenamefont {Hirata}, \citenamefont {Wu}, \citenamefont {Mayaffre},
  \citenamefont {Kr{\"a}mer}, \citenamefont {Liang}, \citenamefont {Hardy},
  \citenamefont {Bonn},\ and\ \citenamefont {Julien}}]{Vinograd:2021aa}%
  \BibitemOpen
  \bibfield  {author} {\bibinfo {author} {\bibfnamefont {I.}~\bibnamefont
  {Vinograd}}, \bibinfo {author} {\bibfnamefont {R.}~\bibnamefont {Zhou}},
  \bibinfo {author} {\bibfnamefont {M.}~\bibnamefont {Hirata}}, \bibinfo
  {author} {\bibfnamefont {T.}~\bibnamefont {Wu}}, \bibinfo {author}
  {\bibfnamefont {H.}~\bibnamefont {Mayaffre}}, \bibinfo {author}
  {\bibfnamefont {S.}~\bibnamefont {Kr{\"a}mer}}, \bibinfo {author}
  {\bibfnamefont {R.}~\bibnamefont {Liang}}, \bibinfo {author} {\bibfnamefont
  {W.~N.}\ \bibnamefont {Hardy}}, \bibinfo {author} {\bibfnamefont {D.~A.}\
  \bibnamefont {Bonn}},\ and\ \bibinfo {author} {\bibfnamefont {M.-H.}\
  \bibnamefont {Julien}},\ }\bibfield  {title} {\bibinfo {title} {{Locally
  commensurate charge-density wave with three-unit-cell periodicity in
  YBa$_2$Cu$_3$O$_y$}},\ }\href {https://doi.org/10.1038/s41467-021-23140-w}
  {\bibfield  {journal} {\bibinfo  {journal} {Nature Communications}\ }\textbf
  {\bibinfo {volume} {12}},\ \bibinfo {pages} {3274} (\bibinfo {year}
  {2021}{\natexlab{a}})}\BibitemShut {NoStop}%
\bibitem [{\citenamefont {Jan{\v s}a}\ \emph {et~al.}(2018)\citenamefont
  {Jan{\v s}a}, \citenamefont {Zorko}, \citenamefont {Gomil{\v s}ek},
  \citenamefont {Pregelj}, \citenamefont {Kr{\"a}mer}, \citenamefont {Biner},
  \citenamefont {Biffin}, \citenamefont {R{\"u}egg},\ and\ \citenamefont
  {Klanj{\v s}ek}}]{Jansa:2018aa}%
  \BibitemOpen
  \bibfield  {author} {\bibinfo {author} {\bibfnamefont {N.}~\bibnamefont
  {Jan{\v s}a}}, \bibinfo {author} {\bibfnamefont {A.}~\bibnamefont {Zorko}},
  \bibinfo {author} {\bibfnamefont {M.}~\bibnamefont {Gomil{\v s}ek}}, \bibinfo
  {author} {\bibfnamefont {M.}~\bibnamefont {Pregelj}}, \bibinfo {author}
  {\bibfnamefont {K.~W.}\ \bibnamefont {Kr{\"a}mer}}, \bibinfo {author}
  {\bibfnamefont {D.}~\bibnamefont {Biner}}, \bibinfo {author} {\bibfnamefont
  {A.}~\bibnamefont {Biffin}}, \bibinfo {author} {\bibfnamefont
  {C.}~\bibnamefont {R{\"u}egg}},\ and\ \bibinfo {author} {\bibfnamefont
  {M.}~\bibnamefont {Klanj{\v s}ek}},\ }\bibfield  {title} {\bibinfo {title}
  {{Observation of two types of fractional excitation in the Kitaev honeycomb
  magnet}},\ }\href {https://doi.org/10.1038/s41567-018-0129-5} {\bibfield
  {journal} {\bibinfo  {journal} {Nature Physics}\ }\textbf {\bibinfo {volume}
  {14}},\ \bibinfo {pages} {786} (\bibinfo {year} {2018})}\BibitemShut
  {NoStop}%
\bibitem [{\citenamefont {Ishida}\ \emph {et~al.}(2020)\citenamefont {Ishida},
  \citenamefont {Manago}, \citenamefont {Kinjo},\ and\ \citenamefont
  {Maeno}}]{Ishida2019}%
  \BibitemOpen
  \bibfield  {author} {\bibinfo {author} {\bibfnamefont {K.}~\bibnamefont
  {Ishida}}, \bibinfo {author} {\bibfnamefont {M.}~\bibnamefont {Manago}},
  \bibinfo {author} {\bibfnamefont {K.}~\bibnamefont {Kinjo}},\ and\ \bibinfo
  {author} {\bibfnamefont {Y.}~\bibnamefont {Maeno}},\ }\bibfield  {title}
  {\bibinfo {title} {Reduction of the {$^{17}$O} knight shift in the
  superconducting state and the heat-up effect by {NMR} pulses on
  {Sr$_2$RuO$_4$}},\ }\href {https://doi.org/10.7566/JPSJ.89.034712} {\bibfield
   {journal} {\bibinfo  {journal} {Journal of the Physical Society of Japan}\
  }\textbf {\bibinfo {volume} {89}},\ \bibinfo {pages} {034712} (\bibinfo
  {year} {2020})}\BibitemShut {NoStop}%
\bibitem [{\citenamefont {Pustogow}\ \emph {et~al.}(2019)\citenamefont
  {Pustogow}, \citenamefont {Luo}, \citenamefont {Chronister}, \citenamefont
  {Su}, \citenamefont {Sokolov}, \citenamefont {Jerzembeck}, \citenamefont
  {Mackenzie}, \citenamefont {Hicks}, \citenamefont {Kikugawa}, \citenamefont
  {Raghu}, \citenamefont {Bauer},\ and\ \citenamefont {Brown}}]{Pustogow2019}%
  \BibitemOpen
  \bibfield  {author} {\bibinfo {author} {\bibfnamefont {A.}~\bibnamefont
  {Pustogow}}, \bibinfo {author} {\bibfnamefont {Y.}~\bibnamefont {Luo}},
  \bibinfo {author} {\bibfnamefont {A.}~\bibnamefont {Chronister}}, \bibinfo
  {author} {\bibfnamefont {Y.-S.}\ \bibnamefont {Su}}, \bibinfo {author}
  {\bibfnamefont {D.~A.}\ \bibnamefont {Sokolov}}, \bibinfo {author}
  {\bibfnamefont {F.}~\bibnamefont {Jerzembeck}}, \bibinfo {author}
  {\bibfnamefont {A.~P.}\ \bibnamefont {Mackenzie}}, \bibinfo {author}
  {\bibfnamefont {C.~W.}\ \bibnamefont {Hicks}}, \bibinfo {author}
  {\bibfnamefont {N.}~\bibnamefont {Kikugawa}}, \bibinfo {author}
  {\bibfnamefont {S.}~\bibnamefont {Raghu}}, \bibinfo {author} {\bibfnamefont
  {E.~D.}\ \bibnamefont {Bauer}},\ and\ \bibinfo {author} {\bibfnamefont
  {S.~E.}\ \bibnamefont {Brown}},\ }\bibfield  {title} {\bibinfo {title}
  {Constraints on the superconducting order parameter in {Sr$_2$RuO$_4$} from
  {Oxygen-17} nuclear magnetic resonance},\ }\href
  {https://doi.org/10.1038/s41586-019-1596-2} {\bibfield  {journal} {\bibinfo
  {journal} {Nature}\ }\textbf {\bibinfo {volume} {574}},\ \bibinfo {pages}
  {72} (\bibinfo {year} {2019})}\BibitemShut {NoStop}%
\bibitem [{\citenamefont {Vinograd}\ \emph
  {et~al.}(2021{\natexlab{b}})\citenamefont {Vinograd}, \citenamefont
  {Edwards}, \citenamefont {Wang}, \citenamefont {Kissikov}, \citenamefont
  {Byland}, \citenamefont {Badger}, \citenamefont {Taufour},\ and\
  \citenamefont {Curro}}]{Vinograd2021}%
  \BibitemOpen
  \bibfield  {author} {\bibinfo {author} {\bibfnamefont {I.}~\bibnamefont
  {Vinograd}}, \bibinfo {author} {\bibfnamefont {S.~P.}\ \bibnamefont
  {Edwards}}, \bibinfo {author} {\bibfnamefont {Z.}~\bibnamefont {Wang}},
  \bibinfo {author} {\bibfnamefont {T.}~\bibnamefont {Kissikov}}, \bibinfo
  {author} {\bibfnamefont {J.~K.}\ \bibnamefont {Byland}}, \bibinfo {author}
  {\bibfnamefont {J.~R.}\ \bibnamefont {Badger}}, \bibinfo {author}
  {\bibfnamefont {V.}~\bibnamefont {Taufour}},\ and\ \bibinfo {author}
  {\bibfnamefont {N.~J.}\ \bibnamefont {Curro}},\ }\bibfield  {title} {\bibinfo
  {title} {{Inhomogeneous Knight shift in vortex cores of superconducting
  FeSe}},\ }\href {https://doi.org/10.1103/PhysRevB.104.014502} {\bibfield
  {journal} {\bibinfo  {journal} {Phys. Rev. B}\ }\textbf {\bibinfo {volume}
  {104}},\ \bibinfo {pages} {014502} (\bibinfo {year}
  {2021}{\natexlab{b}})}\BibitemShut {NoStop}%
\bibitem [{\citenamefont {Mitrovi\ifmmode~\acute{c}\else \'{c}\fi{}}\ \emph
  {et~al.}(2008)\citenamefont {Mitrovi\ifmmode~\acute{c}\else \'{c}\fi{}},
  \citenamefont {Koutroulakis}, \citenamefont {Klanj\ifmmode~\check{s}\else
  \v{s}\fi{}ek}, \citenamefont {Horvati\ifmmode~\acute{c}\else \'{c}\fi{}},
  \citenamefont {Berthier}, \citenamefont {Knebel}, \citenamefont {Lapertot},\
  and\ \citenamefont {Flouquet}}]{Mitrovic2008}%
  \BibitemOpen
  \bibfield  {author} {\bibinfo {author} {\bibfnamefont {V.~F.}\ \bibnamefont
  {Mitrovi\ifmmode~\acute{c}\else \'{c}\fi{}}}, \bibinfo {author}
  {\bibfnamefont {G.}~\bibnamefont {Koutroulakis}}, \bibinfo {author}
  {\bibfnamefont {M.}~\bibnamefont {Klanj\ifmmode~\check{s}\else
  \v{s}\fi{}ek}}, \bibinfo {author} {\bibfnamefont {M.}~\bibnamefont
  {Horvati\ifmmode~\acute{c}\else \'{c}\fi{}}}, \bibinfo {author}
  {\bibfnamefont {C.}~\bibnamefont {Berthier}}, \bibinfo {author}
  {\bibfnamefont {G.}~\bibnamefont {Knebel}}, \bibinfo {author} {\bibfnamefont
  {G.}~\bibnamefont {Lapertot}},\ and\ \bibinfo {author} {\bibfnamefont
  {J.}~\bibnamefont {Flouquet}},\ }\bibfield  {title} {\bibinfo {title}
  {{Comment on ``Texture in the Superconducting Order Parameter of
  {${\mathrm{CeCoIn}}_{5}$} Revealed by Nuclear Magnetic Resonance''}},\ }\href
  {https://doi.org/10.1103/PhysRevLett.101.039701} {\bibfield  {journal}
  {\bibinfo  {journal} {Phys. Rev. Lett.}\ }\textbf {\bibinfo {volume} {101}},\
  \bibinfo {pages} {039701} (\bibinfo {year} {2008})}\BibitemShut {NoStop}%
\bibitem [{\citenamefont {Carr}\ \emph {et~al.}(2022)\citenamefont {Carr},
  \citenamefont {Snider}, \citenamefont {Feldman}, \citenamefont {Ramanathan},
  \citenamefont {Marston},\ and\ \citenamefont {Mitrovi\ifmmode~\acute{c}\else
  \'{c}\fi{}}}]{carr2022signatures}%
  \BibitemOpen
  \bibfield  {author} {\bibinfo {author} {\bibfnamefont {S.}~\bibnamefont
  {Carr}}, \bibinfo {author} {\bibfnamefont {C.}~\bibnamefont {Snider}},
  \bibinfo {author} {\bibfnamefont {D.~E.}\ \bibnamefont {Feldman}}, \bibinfo
  {author} {\bibfnamefont {C.}~\bibnamefont {Ramanathan}}, \bibinfo {author}
  {\bibfnamefont {J.~B.}\ \bibnamefont {Marston}},\ and\ \bibinfo {author}
  {\bibfnamefont {V.~F.}\ \bibnamefont {Mitrovi\ifmmode~\acute{c}\else
  \'{c}\fi{}}},\ }\bibfield  {title} {\bibinfo {title} {Signatures of
  electronic correlations and spin-susceptibility anisotropy in nuclear
  magnetic resonance},\ }\href {https://doi.org/10.1103/PhysRevB.106.L041119}
  {\bibfield  {journal} {\bibinfo  {journal} {Phys. Rev. B}\ }\textbf {\bibinfo
  {volume} {106}},\ \bibinfo {pages} {L041119} (\bibinfo {year}
  {2022})}\BibitemShut {NoStop}%
\bibitem [{\citenamefont {Li}\ \emph {et~al.}(2021)\citenamefont {Li},
  \citenamefont {Hansen}, \citenamefont {Yuan}, \citenamefont
  {Bruschweiler-Li},\ and\ \citenamefont {Br{\"u}schweiler}}]{li2021deep}%
  \BibitemOpen
  \bibfield  {author} {\bibinfo {author} {\bibfnamefont {D.-W.}\ \bibnamefont
  {Li}}, \bibinfo {author} {\bibfnamefont {A.~L.}\ \bibnamefont {Hansen}},
  \bibinfo {author} {\bibfnamefont {C.}~\bibnamefont {Yuan}}, \bibinfo {author}
  {\bibfnamefont {L.}~\bibnamefont {Bruschweiler-Li}},\ and\ \bibinfo {author}
  {\bibfnamefont {R.}~\bibnamefont {Br{\"u}schweiler}},\ }\bibfield  {title}
  {\bibinfo {title} {{DEEP picker is a deep neural network for accurate
  deconvolution of complex two-dimensional NMR spectra}},\ }\href
  {https://doi.org/10.1038/s41467-021-25496-5} {\bibfield  {journal} {\bibinfo
  {journal} {Nature Communications}\ }\textbf {\bibinfo {volume} {12}},\
  \bibinfo {pages} {5229} (\bibinfo {year} {2021})}\BibitemShut {NoStop}%
\bibitem [{\citenamefont {Cordova}\ \emph {et~al.}(2021)\citenamefont
  {Cordova}, \citenamefont {Balodis}, \citenamefont {de~Almeida}, \citenamefont
  {Ceriotti},\ and\ \citenamefont {Emsley}}]{cordova2021bayesian}%
  \BibitemOpen
  \bibfield  {author} {\bibinfo {author} {\bibfnamefont {M.}~\bibnamefont
  {Cordova}}, \bibinfo {author} {\bibfnamefont {M.}~\bibnamefont {Balodis}},
  \bibinfo {author} {\bibfnamefont {B.~S.}\ \bibnamefont {de~Almeida}},
  \bibinfo {author} {\bibfnamefont {M.}~\bibnamefont {Ceriotti}},\ and\
  \bibinfo {author} {\bibfnamefont {L.}~\bibnamefont {Emsley}},\ }\bibfield
  {title} {\bibinfo {title} {Bayesian probabilistic assignment of chemical
  shifts in organic solids},\ }\href {https://doi.org/10.1126/sciadv.abk2341}
  {\bibfield  {journal} {\bibinfo  {journal} {Science Advances}\ }\textbf
  {\bibinfo {volume} {7}},\ \bibinfo {pages} {eabk2341} (\bibinfo {year}
  {2021})}\BibitemShut {NoStop}%
\bibitem [{\citenamefont {Cobas}(2020)}]{cobas2020nmr}%
  \BibitemOpen
  \bibfield  {author} {\bibinfo {author} {\bibfnamefont {C.}~\bibnamefont
  {Cobas}},\ }\bibfield  {title} {\bibinfo {title} {{NMR signal processing,
  prediction, and structure verification with machine learning techniques}},\
  }\href {https://doi.org/https://doi.org/10.1002/mrc.4989} {\bibfield
  {journal} {\bibinfo  {journal} {Magnetic Resonance in Chemistry}\ }\textbf
  {\bibinfo {volume} {58}},\ \bibinfo {pages} {512} (\bibinfo {year}
  {2020})}\BibitemShut {NoStop}%
\bibitem [{\citenamefont {Bratko}(1997)}]{bratko1997machine}%
  \BibitemOpen
  \bibfield  {author} {\bibinfo {author} {\bibfnamefont {I.}~\bibnamefont
  {Bratko}},\ }\bibfield  {title} {\bibinfo {title} {Machine learning: Between
  accuracy and interpretability},\ }in\ \href
  {https://link.springer.com/chapter/10.1007/978-3-7091-2668-4_10} {\emph
  {\bibinfo {booktitle} {Learning, Networks and Statistics}}},\ \bibinfo
  {editor} {edited by\ \bibinfo {editor} {\bibfnamefont {G.}~\bibnamefont
  {Della~Riccia}}, \bibinfo {editor} {\bibfnamefont {H.-J.}\ \bibnamefont
  {Lenz}},\ and\ \bibinfo {editor} {\bibfnamefont {R.}~\bibnamefont {Kruse}}}\
  (\bibinfo  {publisher} {Springer Vienna},\ \bibinfo {address} {Vienna},\
  \bibinfo {year} {1997})\ pp.\ \bibinfo {pages} {163--177}\BibitemShut
  {NoStop}%
\bibitem [{\citenamefont {Carleo}\ \emph {et~al.}(2019)\citenamefont {Carleo},
  \citenamefont {Cirac}, \citenamefont {Cranmer}, \citenamefont {Daudet},
  \citenamefont {Schuld}, \citenamefont {Tishby}, \citenamefont
  {Vogt-Maranto},\ and\ \citenamefont {Zdeborov\'a}}]{carleo2019machine}%
  \BibitemOpen
  \bibfield  {author} {\bibinfo {author} {\bibfnamefont {G.}~\bibnamefont
  {Carleo}}, \bibinfo {author} {\bibfnamefont {I.}~\bibnamefont {Cirac}},
  \bibinfo {author} {\bibfnamefont {K.}~\bibnamefont {Cranmer}}, \bibinfo
  {author} {\bibfnamefont {L.}~\bibnamefont {Daudet}}, \bibinfo {author}
  {\bibfnamefont {M.}~\bibnamefont {Schuld}}, \bibinfo {author} {\bibfnamefont
  {N.}~\bibnamefont {Tishby}}, \bibinfo {author} {\bibfnamefont
  {L.}~\bibnamefont {Vogt-Maranto}},\ and\ \bibinfo {author} {\bibfnamefont
  {L.}~\bibnamefont {Zdeborov\'a}},\ }\bibfield  {title} {\bibinfo {title}
  {Machine learning and the physical sciences},\ }\href
  {https://doi.org/10.1103/RevModPhys.91.045002} {\bibfield  {journal}
  {\bibinfo  {journal} {Rev. Mod. Phys.}\ }\textbf {\bibinfo {volume} {91}},\
  \bibinfo {pages} {045002} (\bibinfo {year} {2019})}\BibitemShut {NoStop}%
\bibitem [{\citenamefont {Jolliffe}\ and\ \citenamefont
  {Cadima}(2016)}]{jolliffe2016principal}%
  \BibitemOpen
  \bibfield  {author} {\bibinfo {author} {\bibfnamefont {I.~T.}\ \bibnamefont
  {Jolliffe}}\ and\ \bibinfo {author} {\bibfnamefont {J.}~\bibnamefont
  {Cadima}},\ }\bibfield  {title} {\bibinfo {title} {Principal component
  analysis: a review and recent developments},\ }\href
  {https://doi.org/10.1098/rsta.2015.0202} {\bibfield  {journal} {\bibinfo
  {journal} {Philosophical Transactions of the Royal Society A: Mathematical,
  Physical and Engineering Sciences}\ }\textbf {\bibinfo {volume} {374}},\
  \bibinfo {pages} {20150202} (\bibinfo {year} {2016})}\BibitemShut {NoStop}%
\bibitem [{\citenamefont {{Kingma}}\ and\ \citenamefont
  {{Welling}}(2013)}]{kingma2013auto}%
  \BibitemOpen
  \bibfield  {author} {\bibinfo {author} {\bibfnamefont {D.~P.}\ \bibnamefont
  {{Kingma}}}\ and\ \bibinfo {author} {\bibfnamefont {M.}~\bibnamefont
  {{Welling}}},\ }\bibfield  {title} {\bibinfo {title} {{Auto-Encoding
  Variational Bayes}},\ }\href {https://arxiv.org/abs/1312.6114} {\bibfield
  {journal} {\bibinfo  {journal} {arXiv e-prints}\ ,\ \bibinfo {eid}
  {arXiv:1312.6114}} (\bibinfo {year} {2013})}\BibitemShut {NoStop}%
\bibitem [{\citenamefont {Torrisi}\ \emph {et~al.}(2020)\citenamefont
  {Torrisi}, \citenamefont {Carbone}, \citenamefont {Rohr}, \citenamefont
  {Montoya}, \citenamefont {Ha}, \citenamefont {Yano}, \citenamefont {Suram},\
  and\ \citenamefont {Hung}}]{torrisi2020random}%
  \BibitemOpen
  \bibfield  {author} {\bibinfo {author} {\bibfnamefont {S.~B.}\ \bibnamefont
  {Torrisi}}, \bibinfo {author} {\bibfnamefont {M.~R.}\ \bibnamefont
  {Carbone}}, \bibinfo {author} {\bibfnamefont {B.~A.}\ \bibnamefont {Rohr}},
  \bibinfo {author} {\bibfnamefont {J.~H.}\ \bibnamefont {Montoya}}, \bibinfo
  {author} {\bibfnamefont {Y.}~\bibnamefont {Ha}}, \bibinfo {author}
  {\bibfnamefont {J.}~\bibnamefont {Yano}}, \bibinfo {author} {\bibfnamefont
  {S.~K.}\ \bibnamefont {Suram}},\ and\ \bibinfo {author} {\bibfnamefont
  {L.}~\bibnamefont {Hung}},\ }\bibfield  {title} {\bibinfo {title} {{Random
  forest machine learning models for interpretable X-ray absorption near-edge
  structure spectrum-property relationships}},\ }\href
  {https://doi.org/10.1038/s41524-020-00376-6} {\bibfield  {journal} {\bibinfo
  {journal} {npj Computational Materials}\ }\textbf {\bibinfo {volume} {6}},\
  \bibinfo {pages} {109} (\bibinfo {year} {2020})}\BibitemShut {NoStop}%
\bibitem [{\citenamefont {Chawla}\ \emph {et~al.}(2002)\citenamefont {Chawla},
  \citenamefont {Bowyer}, \citenamefont {Hall},\ and\ \citenamefont
  {Kegelmeyer}}]{chawla2002smote}%
  \BibitemOpen
  \bibfield  {author} {\bibinfo {author} {\bibfnamefont {N.~V.}\ \bibnamefont
  {Chawla}}, \bibinfo {author} {\bibfnamefont {K.~W.}\ \bibnamefont {Bowyer}},
  \bibinfo {author} {\bibfnamefont {L.~O.}\ \bibnamefont {Hall}},\ and\
  \bibinfo {author} {\bibfnamefont {W.~P.}\ \bibnamefont {Kegelmeyer}},\
  }\bibfield  {title} {\bibinfo {title} {{SMOTE: synthetic minority
  over-sampling technique}},\ }\href {https://doi.org/10.1613/jair.953}
  {\bibfield  {journal} {\bibinfo  {journal} {Journal of Artificial
  Intelligence Research}\ }\textbf {\bibinfo {volume} {16}},\ \bibinfo {pages}
  {321} (\bibinfo {year} {2002})}\BibitemShut {NoStop}%
\bibitem [{\citenamefont {Mavadia}\ \emph {et~al.}(2017)\citenamefont
  {Mavadia}, \citenamefont {Frey}, \citenamefont {Sastrawan}, \citenamefont
  {Dona},\ and\ \citenamefont {Biercuk}}]{Mavadia2017}%
  \BibitemOpen
  \bibfield  {author} {\bibinfo {author} {\bibfnamefont {S.}~\bibnamefont
  {Mavadia}}, \bibinfo {author} {\bibfnamefont {V.}~\bibnamefont {Frey}},
  \bibinfo {author} {\bibfnamefont {J.}~\bibnamefont {Sastrawan}}, \bibinfo
  {author} {\bibfnamefont {S.}~\bibnamefont {Dona}},\ and\ \bibinfo {author}
  {\bibfnamefont {M.~J.}\ \bibnamefont {Biercuk}},\ }\bibfield  {title}
  {\bibinfo {title} {Prediction and real-time compensation of qubit decoherence
  via machine learning},\ }\href {https://doi.org/10.1038/ncomms14106}
  {\bibfield  {journal} {\bibinfo  {journal} {Nature Communications}\ }\textbf
  {\bibinfo {volume} {8}},\ \bibinfo {pages} {14106} (\bibinfo {year}
  {2017})}\BibitemShut {NoStop}%
\bibitem [{\citenamefont {Radovic}\ \emph {et~al.}(2018)\citenamefont
  {Radovic}, \citenamefont {Williams}, \citenamefont {Rousseau}, \citenamefont
  {Kagan}, \citenamefont {Bonacorsi}, \citenamefont {Himmel}, \citenamefont
  {Aurisano}, \citenamefont {Terao},\ and\ \citenamefont
  {Wongjirad}}]{Radovic2018}%
  \BibitemOpen
  \bibfield  {author} {\bibinfo {author} {\bibfnamefont {A.}~\bibnamefont
  {Radovic}}, \bibinfo {author} {\bibfnamefont {M.}~\bibnamefont {Williams}},
  \bibinfo {author} {\bibfnamefont {D.}~\bibnamefont {Rousseau}}, \bibinfo
  {author} {\bibfnamefont {M.}~\bibnamefont {Kagan}}, \bibinfo {author}
  {\bibfnamefont {D.}~\bibnamefont {Bonacorsi}}, \bibinfo {author}
  {\bibfnamefont {A.}~\bibnamefont {Himmel}}, \bibinfo {author} {\bibfnamefont
  {A.}~\bibnamefont {Aurisano}}, \bibinfo {author} {\bibfnamefont
  {K.}~\bibnamefont {Terao}},\ and\ \bibinfo {author} {\bibfnamefont
  {T.}~\bibnamefont {Wongjirad}},\ }\bibfield  {title} {\bibinfo {title}
  {Machine learning at the energy and intensity frontiers of particle
  physics},\ }\href {https://doi.org/10.1038/s41586-018-0361-2} {\bibfield
  {journal} {\bibinfo  {journal} {Nature}\ }\textbf {\bibinfo {volume} {560}},\
  \bibinfo {pages} {41} (\bibinfo {year} {2018})}\BibitemShut {NoStop}%
\bibitem [{\citenamefont {Kutsukake}\ \emph {et~al.}(2020)\citenamefont
  {Kutsukake}, \citenamefont {Nagai}, \citenamefont {Horikawa},\ and\
  \citenamefont {Banba}}]{Kutsukake2020}%
  \BibitemOpen
  \bibfield  {author} {\bibinfo {author} {\bibfnamefont {K.}~\bibnamefont
  {Kutsukake}}, \bibinfo {author} {\bibfnamefont {Y.}~\bibnamefont {Nagai}},
  \bibinfo {author} {\bibfnamefont {T.}~\bibnamefont {Horikawa}},\ and\
  \bibinfo {author} {\bibfnamefont {H.}~\bibnamefont {Banba}},\ }\bibfield
  {title} {\bibinfo {title} {Real-time prediction of interstitial oxygen
  concentration in czochralski silicon using machine learning},\ }\href
  {https://doi.org/10.35848/1882-0786/abc6ec} {\bibfield  {journal} {\bibinfo
  {journal} {Applied Physics Express}\ }\textbf {\bibinfo {volume} {13}},\
  \bibinfo {pages} {125502} (\bibinfo {year} {2020})}\BibitemShut {NoStop}%
\bibitem [{\citenamefont {Levitt}(2007)}]{Levitt2007}%
  \BibitemOpen
  \bibfield  {author} {\bibinfo {author} {\bibfnamefont {M.~H.}\ \bibnamefont
  {Levitt}},\ }\bibinfo {title} {{Symmetry-Based Pulse Sequences in Magic-Angle
  Spinning Solid-State NMR}},\ in\ \href
  {https://doi.org/https://doi.org/10.1002/9780470034590.emrstm0551} {\emph
  {\bibinfo {booktitle} {eMagRes}}}\ (\bibinfo  {publisher} {John Wiley \&
  Sons, Ltd},\ \bibinfo {year} {2007})\BibitemShut {NoStop}%
\bibitem [{\citenamefont {Levitt}(2008)}]{Levitt2008}%
  \BibitemOpen
  \bibfield  {author} {\bibinfo {author} {\bibfnamefont {M.~H.}\ \bibnamefont
  {Levitt}},\ }\bibfield  {title} {\bibinfo {title} {{Symmetry in the design of
  NMR multiple-pulse sequences}},\ }\href {https://doi.org/10.1063/1.2831927}
  {\bibfield  {journal} {\bibinfo  {journal} {The Journal of Chemical Physics}\
  }\textbf {\bibinfo {volume} {128}},\ \bibinfo {pages} {052205} (\bibinfo
  {year} {2008})}\BibitemShut {NoStop}%
\bibitem [{\citenamefont {Schwartz}\ \emph {et~al.}(2018)\citenamefont
  {Schwartz}, \citenamefont {Scheuer}, \citenamefont {Tratzmiller},
  \citenamefont {M{\"u}ller}, \citenamefont {Chen}, \citenamefont {Dhand},
  \citenamefont {Wang}, \citenamefont {M{\"u}ller}, \citenamefont {Naydenov},
  \citenamefont {Jelezko},\ and\ \citenamefont {Plenio}}]{Schwartz2018}%
  \BibitemOpen
  \bibfield  {author} {\bibinfo {author} {\bibfnamefont {I.}~\bibnamefont
  {Schwartz}}, \bibinfo {author} {\bibfnamefont {J.}~\bibnamefont {Scheuer}},
  \bibinfo {author} {\bibfnamefont {B.}~\bibnamefont {Tratzmiller}}, \bibinfo
  {author} {\bibfnamefont {S.}~\bibnamefont {M{\"u}ller}}, \bibinfo {author}
  {\bibfnamefont {Q.}~\bibnamefont {Chen}}, \bibinfo {author} {\bibfnamefont
  {I.}~\bibnamefont {Dhand}}, \bibinfo {author} {\bibfnamefont {Z.-Y.}\
  \bibnamefont {Wang}}, \bibinfo {author} {\bibfnamefont {C.}~\bibnamefont
  {M{\"u}ller}}, \bibinfo {author} {\bibfnamefont {B.}~\bibnamefont
  {Naydenov}}, \bibinfo {author} {\bibfnamefont {F.}~\bibnamefont {Jelezko}},\
  and\ \bibinfo {author} {\bibfnamefont {M.~B.}\ \bibnamefont {Plenio}},\
  }\bibfield  {title} {\bibinfo {title} {{Robust optical polarization of
  nuclear spin baths using Hamiltonian engineering of nitrogen-vacancy center
  quantum dynamics}},\ }\href {https://doi.org/10.1126/sciadv.aat8978}
  {\bibfield  {journal} {\bibinfo  {journal} {Science Advances}\ }\textbf
  {\bibinfo {volume} {4}},\ \bibinfo {pages} {eaat8978} (\bibinfo {year}
  {2018})}\BibitemShut {NoStop}%
\bibitem [{\citenamefont {Higgins}\ \emph {et~al.}(2017)\citenamefont
  {Higgins}, \citenamefont {Matthey}, \citenamefont {Pal}, \citenamefont
  {Burgess}, \citenamefont {Glorot}, \citenamefont {Botvinick}, \citenamefont
  {Mohamed},\ and\ \citenamefont {Lerchner}}]{higgins2016beta}%
  \BibitemOpen
  \bibfield  {author} {\bibinfo {author} {\bibfnamefont {I.}~\bibnamefont
  {Higgins}}, \bibinfo {author} {\bibfnamefont {L.}~\bibnamefont {Matthey}},
  \bibinfo {author} {\bibfnamefont {A.}~\bibnamefont {Pal}}, \bibinfo {author}
  {\bibfnamefont {C.}~\bibnamefont {Burgess}}, \bibinfo {author} {\bibfnamefont
  {X.}~\bibnamefont {Glorot}}, \bibinfo {author} {\bibfnamefont
  {M.}~\bibnamefont {Botvinick}}, \bibinfo {author} {\bibfnamefont
  {S.}~\bibnamefont {Mohamed}},\ and\ \bibinfo {author} {\bibfnamefont
  {A.}~\bibnamefont {Lerchner}},\ }\bibfield  {title} {\bibinfo {title}
  {beta-{VAE}: Learning basic visual concepts with a constrained variational
  framework},\ }in\ \href {https://openreview.net/forum?id=Sy2fzU9gl} {\emph
  {\bibinfo {booktitle} {International Conference on Learning
  Representations}}}\ (\bibinfo {year} {2017})\BibitemShut {NoStop}%
\bibitem [{\citenamefont {James}\ \emph {et~al.}(2013)\citenamefont {James},
  \citenamefont {Witten}, \citenamefont {Hastie},\ and\ \citenamefont
  {Tibshirani}}]{james2013introduction}%
  \BibitemOpen
  \bibfield  {author} {\bibinfo {author} {\bibfnamefont {G.}~\bibnamefont
  {James}}, \bibinfo {author} {\bibfnamefont {D.}~\bibnamefont {Witten}},
  \bibinfo {author} {\bibfnamefont {T.}~\bibnamefont {Hastie}},\ and\ \bibinfo
  {author} {\bibfnamefont {R.}~\bibnamefont {Tibshirani}},\ }\href
  {https://doi.org/10.1007/978-1-4614-7138-7_8} {\emph {\bibinfo {title} {An
  Introduction to Statistical Learning: with Applications in R}}}\ (\bibinfo
  {publisher} {Springer New York},\ \bibinfo {address} {New York, NY},\
  \bibinfo {year} {2013})\ pp.\ \bibinfo {pages} {303--335}\BibitemShut
  {NoStop}%
\bibitem [{\citenamefont {Pedregosa}\ \emph {et~al.}(2011)\citenamefont
  {Pedregosa}, \citenamefont {Varoquaux}, \citenamefont {Gramfort},
  \citenamefont {Michel}, \citenamefont {Thirion}, \citenamefont {Grisel},
  \citenamefont {Blondel}, \citenamefont {Prettenhofer}, \citenamefont {Weiss},
  \citenamefont {Dubourg}, \citenamefont {Vanderplas}, \citenamefont {Passos},
  \citenamefont {Cournapeau}, \citenamefont {Brucher}, \citenamefont {Perrot},\
  and\ \citenamefont {{{\'E}}douard Duchesnay}}]{pedregosa2011scikit}%
  \BibitemOpen
  \bibfield  {author} {\bibinfo {author} {\bibfnamefont {F.}~\bibnamefont
  {Pedregosa}}, \bibinfo {author} {\bibfnamefont {G.}~\bibnamefont
  {Varoquaux}}, \bibinfo {author} {\bibfnamefont {A.}~\bibnamefont {Gramfort}},
  \bibinfo {author} {\bibfnamefont {V.}~\bibnamefont {Michel}}, \bibinfo
  {author} {\bibfnamefont {B.}~\bibnamefont {Thirion}}, \bibinfo {author}
  {\bibfnamefont {O.}~\bibnamefont {Grisel}}, \bibinfo {author} {\bibfnamefont
  {M.}~\bibnamefont {Blondel}}, \bibinfo {author} {\bibfnamefont
  {P.}~\bibnamefont {Prettenhofer}}, \bibinfo {author} {\bibfnamefont
  {R.}~\bibnamefont {Weiss}}, \bibinfo {author} {\bibfnamefont
  {V.}~\bibnamefont {Dubourg}}, \bibinfo {author} {\bibfnamefont
  {J.}~\bibnamefont {Vanderplas}}, \bibinfo {author} {\bibfnamefont
  {A.}~\bibnamefont {Passos}}, \bibinfo {author} {\bibfnamefont
  {D.}~\bibnamefont {Cournapeau}}, \bibinfo {author} {\bibfnamefont
  {M.}~\bibnamefont {Brucher}}, \bibinfo {author} {\bibfnamefont
  {M.}~\bibnamefont {Perrot}},\ and\ \bibinfo {author} {\bibnamefont
  {{{\'E}}douard Duchesnay}},\ }\bibfield  {title} {\bibinfo {title}
  {Scikit-learn: Machine learning in python},\ }\href
  {http://jmlr.org/papers/v12/pedregosa11a.html} {\bibfield  {journal}
  {\bibinfo  {journal} {Journal of Machine Learning Research}\ }\textbf
  {\bibinfo {volume} {12}},\ \bibinfo {pages} {2825} (\bibinfo {year}
  {2011})}\BibitemShut {NoStop}%
\end{thebibliography}%

\end{document}